\documentclass[twocolumn,pra,aps,nofootinbib,superscriptaddress]{revtex4-1}
\usepackage[english]{babel}
\usepackage[utf8]{inputenc}
\usepackage{url,psfrag,graphicx,color}
\usepackage{amsmath,amssymb,amsthm}
\usepackage{bm}
\usepackage{hyperref}
\usepackage{physics}
\usepackage{epsfig}
\usepackage{mathptmx}
\usepackage[english]{babel}
\usepackage{amssymb}
\usepackage{amsfonts}
\usepackage{amsmath}
\usepackage{eucal}
\usepackage{graphicx,color,placeins}
\usepackage{epsfig}
\usepackage{tikz}
\usepackage{cancel}
\usepackage{hyperref}
\usepackage{adjustbox}
\usepackage[normalem]{ulem}

\usepackage{makecell}

\def\<{\left<}
\def\>{\right>}
\def\ket|#1>{\left|#1\right>}
\def\bra<#1|{\left<#1\right|}

\def\elem<#1|#2|#3>{\left<#1\right|#2\left|#3\right>}
\def\({\left(}
\def\){\right)}
\def\[{\left]}
\def\]{\right]}
\def\Z{{\mathbb Z}}
\def\C{{\mathbb C}}
\def\R{{\mathbb R}}
\def\N{{\mathbb N}}
\def\beq{\begin{equation}}
\def\eeq{\end{equation}}

\font\numbers=cmss12
\font\upright=cmu10 scaled\magstep1
\def\stroke{\vrule height8pt width0.4pt depth-0.1pt}
\def\topfleck{\vrule height8pt width0.5pt depth-5.9pt}
\def\botfleck{\vrule height2pt width0.5pt depth0.1pt}
\def\Zmath{\vcenter{\hbox{\numbers\rlap{\rlap{Z}\kern
0.8pt\topfleck}\kern 2.2pt
                   \rlap Z\kern 6pt\botfleck\kern 1pt}}}
\def\Qmath{\vcenter{\hbox{\upright\rlap{\rlap{Q}\kern
                   3.8pt\stroke}\phantom{Q}}}}
\def\Nmath{\vcenter{\hbox{\upright\rlap{I}\kern 1.7pt N}}}
\def\Cmath{\vcenter{\hbox{\upright\rlap{\rlap{C}\kern
                   3.8pt\stroke}\phantom{C}}}}
\def\Rmath{\vcenter{\hbox{\upright\rlap{I}\kern 1.7pt R}}}
\def\Hmath{\vcenter{\hbox{\upright\rlap{I}\kern 1.7pt H}}}
\def\Amath{\vcenter{\hbox{\upright\rlap{I}\kern 1.7pt A}}}
\def\Z{\ifmmode\Zmath\else$\Zmath$\fi}
\def\Q{\ifmmode\Qmath\else$\Qmath$\fi}
\def\N{\ifmmode\Nmath\else$\Nmath$\fi}
\def\C{\ifmmode\Cmath\else$\Cmath$\fi}
\def\R{\ifmmode\Rmath\else$\Rmath$\fi}

\def\ket|#1>{| #1 \rangle}
\def\bra<#1|{\langle #1 |}
\def\<{\langle}
\def\>{\rangle}
\def\{{\lbrace}
\def\}{\rbrace}
\def\({\left(}
\def\){\right)}
\def\[{\left[}
\def\]{\right]}

\def\be{\begin{equation}}
\def\ee{\end{equation}}
\def\bea{\begin{eqnarray}}
\def\eea{\end{eqnarray}}

\def\ket|#1>{| #1 \rangle}
\def\bra<#1|{\langle #1 |}
\def\<{\langle}
\def\>{\rangle}
\def\{{\lbrace}
\def\}{\rbrace}
\def\({\left(}
\def\){\right)}

\def\beq{\begin{equation}}
\def\eeq{\end{equation}}
\def\barray{\begin{eqnarray}}
\def\earray{\end{eqnarray}}

\begin{document}

\title{Bulk-edge correspondence in the Haldane phase of the 
bilinear-biquadratic spin-1 Hamiltonian}

\author{Sudipto Singha Roy}
\affiliation{Instituto de Física Teórica, UAM-CSIC, Universidad
  Aut{\'o}noma de Madrid, Cantoblanco, Madrid, Spain}

\author{Silvia N. Santalla}
\affiliation{Dept. de Física and Grupo Interdisciplinar de Sistemas
  Complejos (GISC), Universidad Carlos III de Madrid, Spain}

\author{Javier Rodríguez-Laguna}
\affiliation{Dept. de  Física Fundamental, UNED, Madrid, Spain}

\author{Germán Sierra}
\affiliation{Instituto de Física Teórica, UAM-CSIC, Universidad
  Aut{\'o}noma de Madrid, Cantoblanco, Madrid, Spain}

\date{\today}

\begin{abstract}
The Haldane phase is the prototype of symmetry protected topological (SPT) phases of spin chain systems. It can be protected by several symmetries having in common the degeneracy of the entanglement spectrum.  Here we explore in depth this degeneracy for the  spin-1 AKLT and bilinear-biquadratic Hamiltonians and show the emergence of a bulk-edge correspondence that relates the low energy properties of the entanglement Hamiltonian of a periodic chain and that of the physical Hamiltonian of an open chain.  We find that the entanglement spectrum can be described in terms of two spins-1/2 behaving as the effective spins at the end of the open chain.   In the case of non-contiguous partitions, we find that the entanglement Hamiltonian is given by the spin-1/2 Heisenberg Hamiltonian, which suggests a relationship between SPT phases and conformal field theory.  We finally investigate the string order parameter and the relation with the bulk-edge correspondence.
\end{abstract}
\maketitle

\section{Introduction}

In a quantum many-body system,  distribution of quantum correlation in its  subparts often provides many insightful details of the underlying physical phenomena. In the past years,  there has been a plethora of works where bipartite quantum entanglement is considered to be an efficient probe to characterize quantum properties of  strongly correlated systems \cite{entanglement_many_body1,entanglement_many_body2,entanglement_many_body3,entanglement_many_body4,entanglement_many_body5,entanglement_many_body6,entanglement_many_body7}. In particular,  entanglement entropy quantifies the amount of entanglement present between a part of the system and its complement and  allows us to distinguish quantum phases that do not break any symmetry and hence can not be characterized using any local order parameter, e.g., topological states of matter \cite{topo_entropy1,topo_entropy2,topo_entropy3,topo_entropy4,topo_entropy5,entanglement_many_body6},  quantum spin liquids \cite{ent_entropy_spin_liquid1,ent_entropy_spin_liquid2,ent_entropy_spin_liquid3}, etc. Additionally, the area law for the entanglement entropy of low-energy states of local Hamiltonians has provided a new framework to describe quantum many-body systems in a more efficient way by exploiting tensor network theory \cite{entanglement_many_body4,entanglement_many_body5, ent_entropy_area_law,tensor_network_bunch1,tensor_network_bunch2,tensor_network_bunch3}.

The von Neumann entanglement entropy summarizes the information about the reduced density matrix for a part of the system into a single number. However, there are scenarios where the distribution of the eigenvalues of the reduced system, commonly known as {\it entanglement spectrum} (ES), can unveil more refined information than the entanglement entropy alone. This has attracted much interest in the community, as several studies have been  reported situations where the ES has been  very insightful. The list includes quantum criticality \cite{ent_spectrum_ref0,ent_spectrum_ref1,ent_spectrum_ref2,ent_spectrum_ref3,ent_spectrum_ref4}, symmetry protected phases of matter \cite{ent_spectrum_ref5,ent_spectrum_ref6,ent_spectrum_ref7,ent_spectrum_ref8,ent_spectrum_ref9},  many-body localization \cite{ent_spectrum_ref10,ent_spectrum_ref11,ent_spectrum_ref12}, etc.  As an example, it has been shown that the degeneracy in the reduced state of the spin-1 chain reflects the non-trivial {\em topological} nature of the Haldane phase of the model \cite{ent_spectrum_ref7}. The degeneracy is protected by the same set of symmetries which protects the Haldane phase. Very recently, a scheme to measure the entanglement spectrum using the IBM quantum computing interface has been demonstrated \cite{ent_spectrum_ref13}. Even further information can be provided by considering the reduced density matrix as a thermal state associated to a certain entanglement Hamiltonian, thus providing a direct route to relate the properties of the subsystem to the physical Hamiltonian of the system. For instance, physics at the edge of topologically ordered phases, e.g., fractional quantum Hall state \cite{ent_spectrum_ref5} and of symmetry-protected topological states of matter \cite{ent_ham_ref2,ent_ham_ref3,ent_ham_ref4,ent_ham_ref5,ent_ham_ref6,ent_ham_ref7} has been found to be imprinted in the entanglement Hamiltonian of a part of the system.  The properties of the entanglement Hamiltonian in conformally invariant systems have also been extensively explored in many recent works  \cite{ent_ham_conformal1,ent_ham_conformal2,ent_ham_conformal3,ent_ham_conformal4,ent_ham_conformal5,ent_ham_conformal6}.  For instance, in Ref.\cite{ent_ham_conformal2} it is shown that for two dimension (2D) CFT, the universal part of the spectrum of the entanglement Hamiltonian is always that of a boundary CFT with appropriate boundary conditions.

In most of the above cases, the entanglement spectrum and the entanglement Hamiltonian have been obtained for contiguous bipartitions of the state. Derivation of entanglement Hamiltonians for non-contiguous blocks of the system is a relatively hard albeit interesting case to explore. In this work, we derive the entanglement Hamiltonian for a topological state built on a spin-1 system for both contiguous and non-contiguous bipartitions of the state. We consider the periodic Affleck-Kennedy-Lieb-Tasaki (AKLT) state \cite{AKLT1} in 1D, which can be obtained as the unique ground state (GS) of the periodic spin-1 bilinear-biquadratic Heisenberg (BBH) Hamiltonian at the AKLT point within the Haldane phase \cite{AKLT2}. Then, we compute the entanglement spectrum for contiguous as well as non-contiguous bipartitions by employing the matrix product states formalism, which provides a more direct route than previous approaches \cite{korepin1,Hirano,korepin2}. We note that in both the cases, the low-energy component of the entanglement Hamiltonian can be approximated as an interacting Heisenberg Hamiltonian between the effective spin-1/2 particles at the boundaries of the blocks. Moreover, for the contiguous case, the low-energy properties of the entanglement Hamiltonian can be related to the edge properties of the associated physical Hamiltonian with open boundaries. This suggests the emergence of a bulk-edge correspondence in the model, similar to the one reported in earlier works \cite{ent_spectrum_ref5,ent_ham_ref2,ent_ham_ref3,ent_ham_ref4,ent_ham_ref5,ent_ham_ref6, ent_ham_ref7}.
In this article, we explore this relation in more depth and report that this bulk-edge correspondence persists even at other points in the Haldane phase, as it is manifested by the high overlap between the low-energy eigenstates of the entanglement and physical Hamiltonians, the identical scaling of their respective energy gaps and the same degeneracy structure of the energy eigenvalues corresponding to the low-energy part of the spectrum. Additionally, we show an intriguing relation between the scaling of both the energy gaps and that of the string order parameter (SOP)  \cite{SOP} in the vicinity of the AKLT point.  As it is well known, the SOP provides a hidden order parameter for the Haldane phase.  In particular, the scaling of SOP with system size remains akin to that of the energy gaps,  which eventually results in a  correlation length  identical to the characteristic length scale associated with the BBH model.
 
This article is organized as follows. In Sec.\ref{Sec1}, we discuss the matrix product state representation of the AKLT state and the form of the corresponding transfer matrix which will be useful to compute the relevant quantities that we wish to explore. In Sec. \ref{Sec2}, we present the analytical derivation of the ES of contiguous and non-contiguous bipartitions of the AKLT state. This is followed by the derivation and characterization of the corresponding entanglement Hamiltonian in both cases. On the other hand, Sec. \ref{Sec3} is devoted to the characterization of the bulk-edge correspondence that emerges between the low-energy spectra of the entanglement Hamiltonian and the corresponding physical Hamiltonian, which is shown to persist even at other points in the Haldane phase. Finally, in Sec. \ref{Sec4} we summarize our conclusions and discuss possible lines of future work.


\section{Preliminaries}
\label{Sec1}

We start with the matrix product states (MPS) representation of the periodic AKLT state, given by
\begin{eqnarray}
|\psi\rangle_{\text{AKLT}}&=&\sum_{i_1 i_2 \dots i_N} \text{Tr}(A_{i_1} A_{i_2}\dots A_{i_N}) |i_1 i_2\dots i_N\rangle,
\label{eqn:AKLT_MPS}
\end{eqnarray}
where the index `$k$' in $A_k$ ($k \in 0, 1, 2$) labels the standard spin-1 basis states in each physical site, with $A_0=\sqrt{\frac{2}{3}}\sigma^+$, $A_1=-\sqrt{\frac{1}{3}}\sigma _z$, $A_2=-\sqrt{\frac{2}{3}}\sigma^-$, $\sigma_z$ is the $z$-component of the Pauli spin vector and $\sigma^+(\sigma^-)$ is spin raising (lowering) operator.  Such a representation of the AKLT state guarantees its canonical form, $\sum_k A_k A_k^ {\dagger}=\mathbb{I}$ (right canonical), and $\sum_k A_k^\dagger A_k=\mathbb{I}$ (left canonical) \cite{MPS_Garcia}. For the sake of efficiency, it is customary to define a transfer matrix as 
\begin{eqnarray}
E=\sum_k A_k \otimes A^*_k.
\label{eqn:Transfer_matrix}
\end{eqnarray}
 The transfer matrix $E$ is diagonalizable and can be expressed in terms of its left and right eigenvectors as follows $E=\sum_{i=0}^{D^2-1} \gamma_i |R_i\rangle\langle L_i|$. For the above choices of the $A_k$ matrices for AKLT state, the set of vectors $\{|L_i\rangle\}_{i=0}^{D^2-1}$ and $ \{|R_i\rangle\}_{i=0}^{D^2-1}$ present the following form in the computational basis,

\begin{eqnarray}
|R_0\rangle=|L_0\rangle&=&\frac{1}{\sqrt{2}} (|00\rangle +|11\rangle),\nonumber\\
|R_1\rangle=|L_1\rangle&=&\frac{1}{\sqrt{2}} (|00\rangle -|11\rangle),\nonumber\\
|R_2\rangle=|L_2\rangle&=&|01\rangle,\nonumber\\
|R_3\rangle=|L_3\rangle&=&|10\rangle,
\label{eqn:transfer_matrix_vec}
\end{eqnarray}
with $\gamma_0=1, \gamma_1=\gamma_2=\gamma_3=\gamma=-\frac{1}{3}$. 

Along with this, here we introduce the quantum many-body Hamiltonian we will use in our work, known as spin-1 bilinear-biquadratic Heisenberg (BBH) Hamiltonian, 
\begin{eqnarray}
H_{BBH}=\sum_i \cos(\theta)\, \vec{S}_i\cdot\vec{S}_{i+1} + \sin(\theta)\, (\vec{S}_i\cdot\vec{S}_{i+1})^2,
\label{eqn:BBH}
\end{eqnarray}
where $S^k$ are spin-1 operators, $k\in x,y,z$.  The AKLT state described above appears at $\theta_{AKLT}=\arctan(\frac{1}{3})$.


\section{Derivation of Entanglement Hamiltonian}
\label{Sec2}

Let us evaluate the entanglement Hamiltonian for both contiguous and non-contiguous subsystems of a periodic AKLT state. In general, the entanglement Hamiltonian $\mathcal{H}_E$ is defined in terms of the reduced density matrix $\rho$ of a quantum state,

\begin{eqnarray}
\mathcal{H}_E=-\ln(\rho). 
\label{eqn:Ent_Ham}
\end{eqnarray}
The above representation often provides an efficient way to unveil the information in the reduced subsystem of the system \cite{ent_spectrum_ref5,ent_ham_ref2,ent_ham_ref3,ent_ham_ref4,ent_ham_ref5,ent_ham_ref6,
ent_ham_ref7,ent_ham_conformal1,ent_ham_conformal2,ent_ham_conformal3,ent_ham_conformal4,
ent_ham_conformal5,ent_ham_conformal6}. In particular, for experimental purposes where direct experimental measurement of the entanglement spectrum requires full state tomography, it may be more feasible to extract the entanglement properties by engineering the entanglement Hamiltonian \cite{ent_ham_exp1}. However, the derivation of the exact form of the entanglement Hamiltonian for a generic quantum many-body state is not straightforward. In our case, we first obtain the exact mathematical form of the reduced density matrices for different choices of the subsystems of the AKLT state and use Eq. (\ref{eqn:Ent_Ham}) to derive the corresponding entanglement Hamiltonian. The detailed methodology is discussed in the following paragraphs.

\subsection{Contiguous bipartition}
\label{Contiguous bipartition}
Let us consider the entanglement Hamiltonian for an $l$-site contiguous subsystem  of the periodic AKLT state.  We decompose the MPS representation of the AKLT state given in Eq. (\ref{eqn:AKLT_MPS}) in the   $A:B$ bipartition as follows
\begin{eqnarray}
|\psi\rangle_{\text{AKLT}}=&&\sum_{\alpha \beta} |\phi^l_{\alpha\beta}\rangle |\phi_{ \beta\alpha}^{N-l}\rangle,
\label{eqn:AKLT_contiguous}
\end{eqnarray}
such that sites $1,2\dots l \in A$ and $l+1,l+2\dots N \in B$ and $\alpha, \beta \in \{0, 1\}$. The intermediate steps to arrive at such decomposition from Eq.  (\ref{eqn:AKLT_MPS}) and the exact mathematical forms of the states $|\phi_{\alpha \beta}^l\rangle$ are given in Appendix \ref{AppendixA}. Though such  decomposition is akin to the Schmidt decomposition in the bipartition $l: N-l$, in general,  $|\phi_{\alpha\beta}^l\rangle$ (or $|\phi_{\alpha\beta}^{N-l}\rangle$) are not mutually orthogonal. For instance, for the AKLT state, the states  $|\phi_{00}^l \rangle$ and $|\phi^l_{11}\rangle$ have non-zero overlap,  which decays with block size $l$ as  $\langle\phi_{00}^l|\phi^l_{11}\rangle$=$\langle \phi_{11}^l |\phi^l_{00}\rangle$=$\gamma^l$. In that case, from the set of states $|\phi_{\alpha \beta}^l\rangle$ one needs to construct an orthogonal basis  $\{|\phi^l_{+}\rangle, |\phi^l_{-}\rangle, |\phi^l_{01}\rangle, |\phi^l_{10}\rangle\}$, where  $|\phi^l_{\pm}\rangle=\frac{1}{\sqrt{2}}(|\phi_{00}^l\rangle\pm |\phi_{11}^l\rangle)$.  We discuss the procedure in detail in Appendix \ref{AppendixA}.

\begin{figure}[t]
\includegraphics[width=8.5cm]{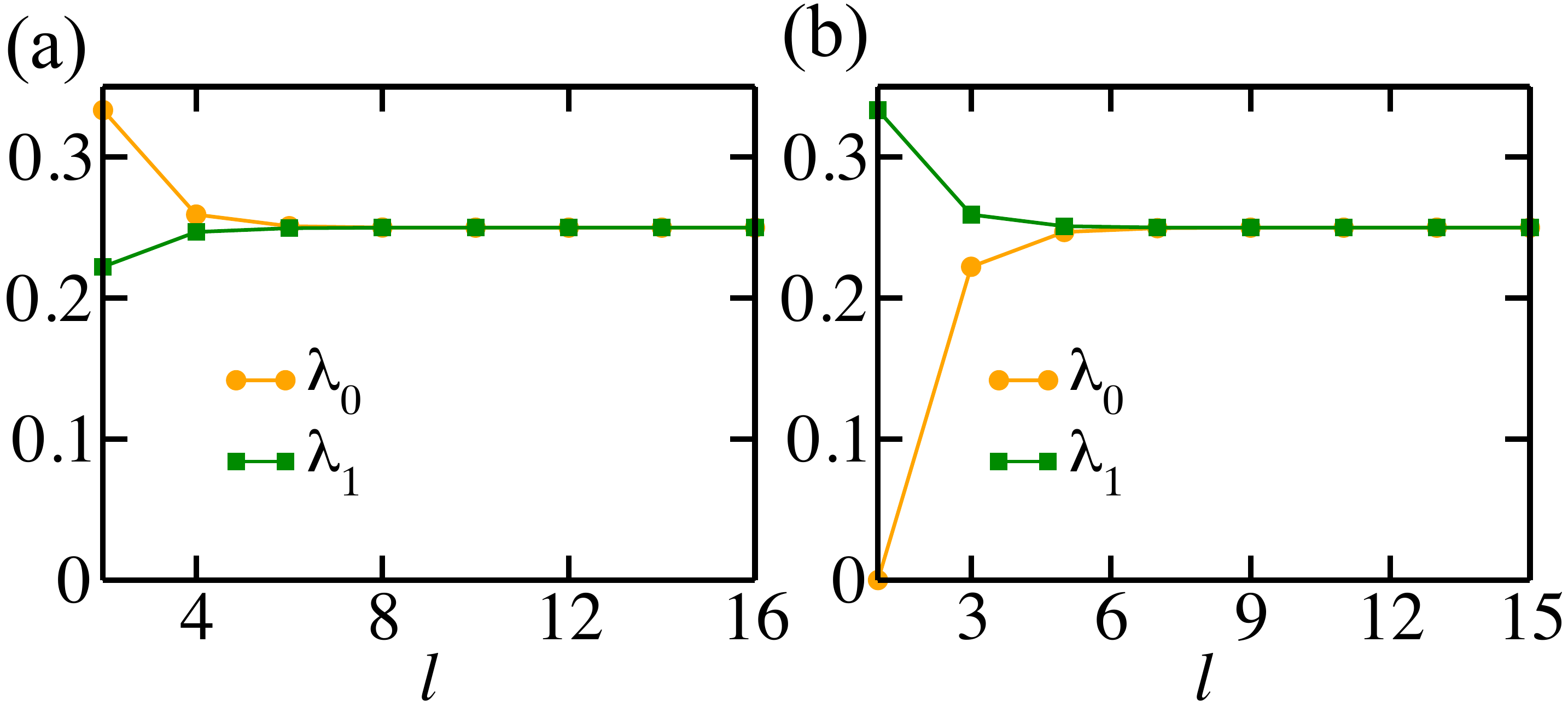}
\caption{Plot of the eigenvalues $\lambda_i$ of the reduced density matrix $\rho$ derived from the contiguous bipartition of the periodic AKLT state as expressed in Eq. (\ref{eqn:lambda_contiguous}). In both panels, we consider the total size of the system $N=50$. In panel (a) we plot the $\lambda_i$'s for even values of $l$ and we can see $\lambda_0\geq\lambda_1$ for all $l$. Whereas, in (b), we plot the $\lambda_i$'s for odd values of $l$ and we get $\lambda_1\geq\lambda_0$ for all $l$. For large $l$,  all the $\lambda_i$'s converge to $\frac{1}{4}$.}
\label{fig:lambda_contiguous}
\end{figure}

Now if we trace out subsystem $B$, the exact form of the $l$-site reduced density matrix $\rho$  in its orthonormal eigenbasis is given by (see Appendix \ref{AppendixA} for a detailed derivation)
\begin{eqnarray}
\rho=\lambda_0|\phi^l_+\rangle \langle \phi^l_+| + \lambda_1|\phi^l_- \rangle\langle \phi^l_-|+\lambda_2|\phi^l_{01}\rangle\langle \phi^l_{01}|+\lambda_3|\phi^l_{10}\rangle \langle \phi^l_{10}|.\nonumber\\
\end{eqnarray}
The eigenvalues $\lambda_i$ for $N\gg1$ can be expressed as

\begin{eqnarray}
\lambda_0&=&\frac{{(1+3\gamma^l) (1+3\gamma^{N-l})}}{4},\nonumber\\
\lambda_1=\lambda_2=\lambda_3&=&\frac{{(1-\gamma^l) (1-\gamma^{N-l})}}{4}.\nonumber\\
\label{eqn:lambda_contiguous}
\end{eqnarray}
The above eigenvalues remain invariant under the exchange of subsystems, i.e., $l$ by $N-l$. Now if we keep total system size $N$  even, for $l$ even, we have $\lambda_0> \lambda_1= \lambda_2= \lambda_3$. Therefore, the spectrum of $\rho$ is comprised of two distinct eigenvalues, the lowest one is triply degenerate and the corresponding  eigenspace is spanned by the basis states  $\{|\phi^l_-\rangle, |\phi^l_{01}\rangle, |\phi^l_{10}\rangle\}$. On the other hand, the highest one is unique and the corresponding eigenstate is given by  $|\phi^l_+\rangle$. Similarly, for odd $l$, we have the opposite ordering of the eigenstates, i,e, in that case the lowest eigenvalue state is unique but the highest eigenvalue is triply degenerate. However, for moderately large values of $l$, this distinction vanishes and all the eigenvalues becomes degenerate and converge to $\frac{1}{4}$. This degeneracy is a signature of the symmetry protected topological (SPT) nature of the AKLT state \cite{ent_spectrum_ref8,SPT_AKLT}. Fig. \ref{fig:lambda_contiguous} depicts the behavior of $\lambda_0$ and $\lambda_1$ for even and odd values of $l$.

Once the exact form of the reduced density matrix $\rho$ has been derived, the expression of the entanglement Hamiltonian can be obtained using Eq. (\ref{eqn:Ent_Ham}). We propose

\begin{eqnarray}
\mathcal{H}_E =-\ln \rho= \varepsilon_0 + J^{(1)}_E\vec{\sigma_1} . \vec{\sigma_2}, 
\label{eqn:EH_contiguous}
\end{eqnarray}
where $\varepsilon_0$ is a constant, $J^{(1)}_E$ is a function of $l$ and $\vec{\sigma}$ is the spin-1/2 vector operator.
This suggests

\begin{eqnarray}
- \log \lambda_0 &=& \varepsilon_0 - 3  J^{(1)}_E,\nonumber\\
- \log \lambda_k &=& \varepsilon_0 +  J^{(1)}_E,
\label{eqn:eigen_ent_Ham}
\end{eqnarray}
with $k\in \{1, 2, 3\}$. At a moderately large value of $l$, $\varepsilon_0 \approx 2\log2+O(\gamma^{2l})$, and $J^{(1)}_E\approx \gamma^l$, which is consistent with the behavior of $\lambda_i$ in this limit. Hence, the  entanglement Hamiltonian remains non-trivial as long as the size of the subsystems remains moderately low such that $J^{(1)}_E$ is non-negligible. Notice that the even-odd dependence that we discussed earlier comes from the fact that $J^{(1)}_E\approx \gamma^l$ with $\gamma=-\frac{1}{3}$. Additionally, the constants $\varepsilon_0$ and $J^{(1)}_E$ can now be interpreted as the GS entanglement energy and entanglement coupling respectively. We note that for even $l$,  $|\phi^l_+\rangle$ becomes the unique GS of the entanglement Hamiltonian and the excited subspace is spanned by the triplet $\{|\phi^l_-\rangle, |\phi^l_{01}\rangle, |\phi^l_{10}\rangle\}$. 
At this stage, in order to unveil characteristic features of the entanglement Hamiltonian, we perform a local transformation of the auxiliary basis, denoted by the index $\alpha$ and $\beta$ in $|\phi_{\alpha\beta}^l\rangle$ (see Appendix\ref{AppendixA}), such that
\begin{figure}[h]
\includegraphics[width=8cm]{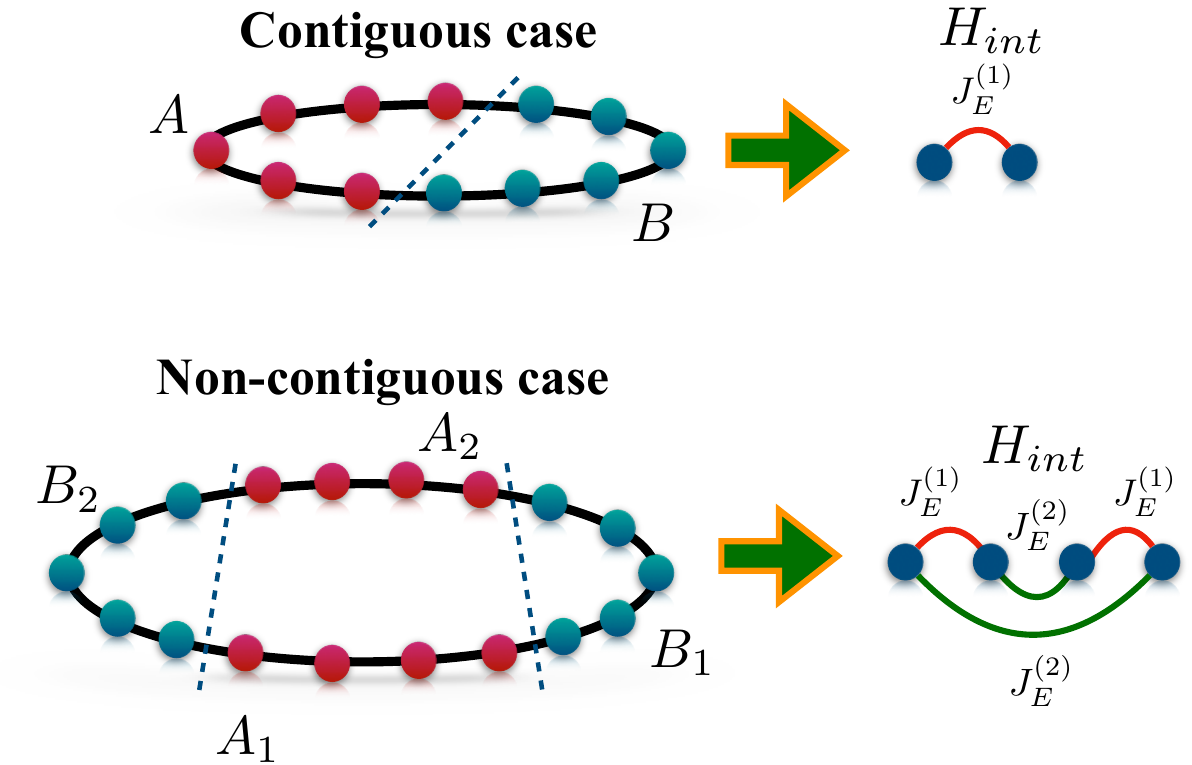}
\caption{Schematic of relation between parts of the AKLT state and the corresponding entanglement Hamiltonian. The entanglement Hamiltonian for a part of the state is similar to the Heisenberg Hamiltonian between the effective spin-1/2 particles at the boundaries of the blocks. The green line in $H_{int}$ represents couplings between inter-block effective spin-1/2 particles ($J_E^{(2)}$), while the red lines represent the coupling between intra-block spins ($J_E^{(1)}$).}
\label{fig:schematic_first}
\end{figure}, 
the eigenbasis of $\mathcal{H}_E$ under this transformation changes as
\begin{eqnarray}
|\phi^l_+\rangle=\frac{(|\phi_{00}^l\rangle+|\phi_{11}^l\rangle)}{\sqrt{2}}&\rightarrow&|\tilde{\phi}^l_+\rangle=\frac{(|\tilde{\phi}^l_{01}\rangle-|\tilde{\phi}^l_{10}\rangle)}{\sqrt{2}},\nonumber\\
|\phi^l_-\rangle=\frac{(|\phi_{00}^l\rangle-|\phi_{11}^l\rangle)}{\sqrt{2}}&\rightarrow&|\tilde{\phi}^l_-\rangle=\frac{(|\tilde{\phi}^l_{01}\rangle+|\tilde{\phi}^l_{10}\rangle)}{\sqrt{2}},\nonumber\\
|\phi^l_{01}\rangle&\rightarrow&-|\tilde{\phi}^l_{00}\rangle,\nonumber\\
|\phi^l_{10}\rangle&\rightarrow&|\tilde{\phi}^l_{11}\rangle.
\end{eqnarray}
Interestingly, a nice interpretation of the entanglement Hamiltonian emerges from the above derivation. In particular, $H_{int}=\vec{\sigma}_1\cdot \vec{\sigma}_2$ stands for the interacting Heisenberg Hamiltonian between the effective spin-1/2 particles at the end of the block (see Fig. \ref{fig:schematic_first}), and the corresponding GS
$|\psi\rangle_G=|\tilde{\phi}_{+}^l\rangle=\frac{|\tilde{\phi}^l_{01}\rangle-|\tilde{\phi}^l_{10}\rangle}{\sqrt{2}}$ can be regarded as a {\em singlet}-state formed in the auxiliary basis $\{|\tilde{\alpha}\rangle, |\tilde{\beta}\rangle\}$ (see Appendix \ref{AppendixA}). Similarly, the excited states of the interaction  Hamiltonian $H_{int}$  ($|\tilde{\phi}^l_-\rangle, |\tilde{\phi}^l_{00}\rangle, |\tilde{\phi}^l_{11}\rangle$) are an analogue to the {\em spin-triplet} state of the Heisenberg Hamiltonian. This interpretation is part of the edge-bulk correspondence that we shall discuss in more detail later on in Sec. \ref{Sec3}.

\subsection{Non-contiguous bipartitions}

Next, let us consider the entanglement Hamiltonian for non-contiguous subsystems of the AKLT model. In particular, we will decompose the AKLT state into the following bipartition,  $A_1: B_1: A_2: B_2$, such that

\begin{eqnarray}
&&[1, 2,\dots,  l_{A_1}] \in A_1,\nonumber\\
&&[l_{A_1}+1, l_{A_1}+2,\dots,   l_{A_1}+l_{B_1}] \in B_1, \nonumber\\
&&[l_{A_1}+l_{B_1}+1, l_{A_1}+l_{B_1}+2,\dots, l_{A_1}+l_{B_1}+l_{A_2}] \in  A_2, \nonumber\\
&&[l_{A_1}+l_{B_1}+l_{A_2}+1, \dots,   l_{A_1}+l_{B_1}+l_{A_2}+l_{B_2}]\in B_2,
\end{eqnarray} 
where $l_{A_1}+l_{B_1}+l_{A_2}+l_{B_2}=N$, $A_1\cup A_2=A$, and  $B_1\cup B_2=B$. 
To find the reduced density matrix in this case, we start with a similar decomposition of the AKLT state as given in Eq. (\ref{eqn:AKLT_contiguous})
\begin{eqnarray}
|\psi\rangle_{\text{AKLT}}&=&\sum_{\alpha \beta}|\phi_{\alpha \beta}^{l_{A_1}+l_{B_1}}\rangle  |\phi_{\beta \alpha}^{l_{A_2}+l_{B_2}}\rangle.
\end{eqnarray}
Further decomposition of the basis $|\phi_{\alpha \beta}^{l_{A_1}+l_{B_1}}\rangle$ and $|\phi_{\beta \alpha}^{l_{A_2}+l_{B_2}}\rangle$ yields
\begin{eqnarray}
|\psi\rangle_{\text{AKLT}}&=& \sum_{\alpha  \gamma  \beta \delta}  |\phi_{\alpha \gamma}^{l_{A_1}}\rangle   |\phi_{\gamma \beta}^{l_{B_1}}\rangle |\phi^{l_{A_2}}_{\beta \delta}\rangle|\phi^{l_{B_2}}_{\delta \alpha}\rangle,\nonumber\\
&=& \sum_{\alpha  \gamma  \beta \delta} \underbrace{ |\phi_{\alpha \gamma}^{l_{A_1}}\rangle   |\phi^{l_{A_2}}_{\beta \delta}}_{A}\rangle \underbrace{|\phi _{\gamma \beta}^{l_{B_1}}\rangle |\phi^{l_{B_2}}_{\delta \alpha}\rangle}_{B}.
\end{eqnarray}
The above representation of the AKLT state now becomes very similar to the contiguous case. Therefore, in order to obtain the reduced density matrix for the subsystem $A=A_1\cup A_2$, we can employ the same techniques as before. However, note that, since the indices $\alpha, \beta, \gamma, \delta$ can take two values, 0 and 1, the reduced state is spanned by sixteen basis states. Similarly to the contiguous case, we find that most of the basis states $|\phi_{\alpha \gamma}^{l_{A_1}}\rangle  |\phi^{l_{A_2}}_{\beta \delta}\rangle$ are not mutually orthogonal. The procedure of construction of mutually orthogonal basis states from these non-orthogonal states is much more complex than the previous case. We discuss the details in Appendix \ref{AppendixA}. Let us stress that the reduced density matrix obtained in this case depends on the parameters $l_{A_1}$, $l_{B_1}$, $l_{A_2}$ and $l_{B_2}$.

\begin{figure}[t]
\includegraphics[width=8.5cm]{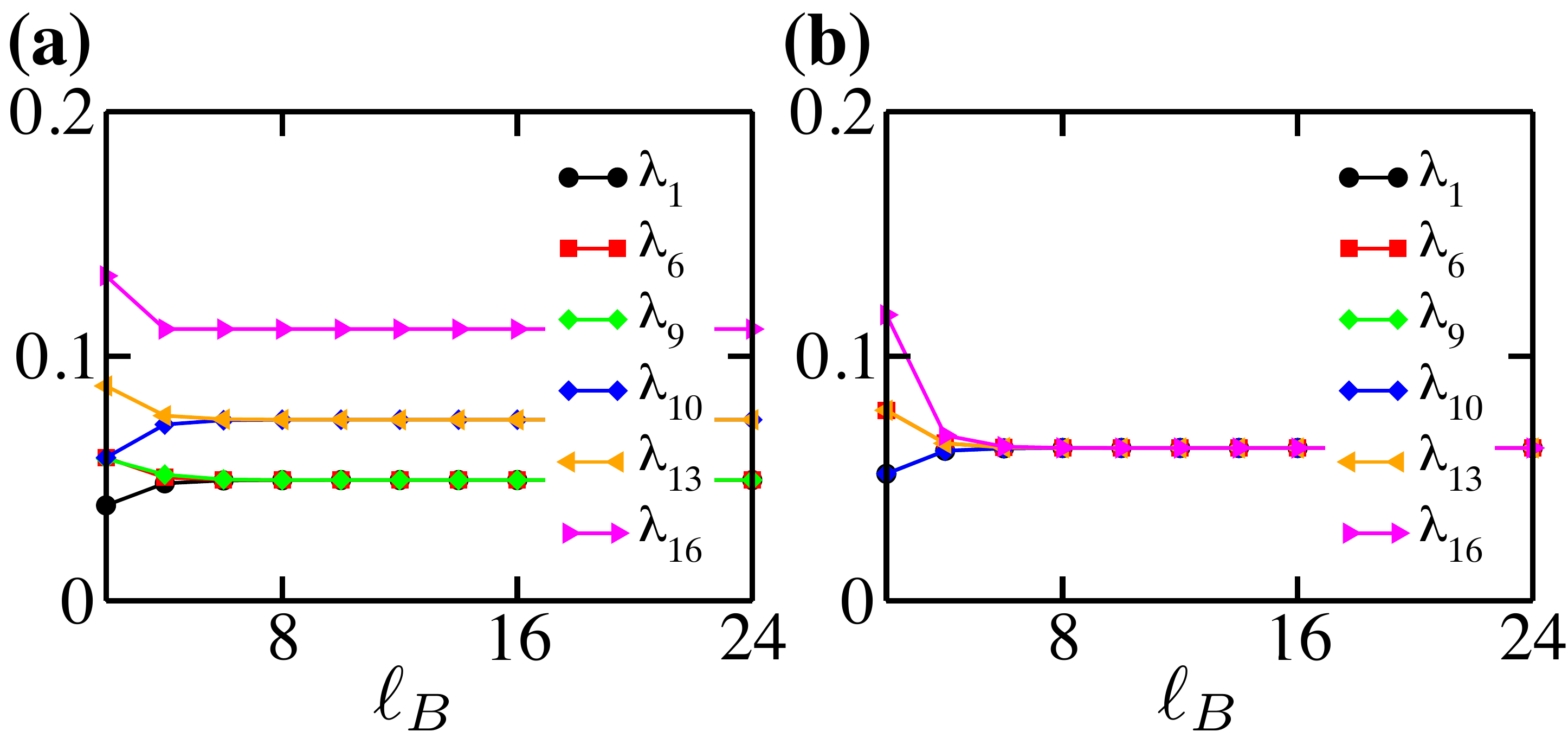}
\caption{Eigenvalues ($\lambda_1$, $\lambda_6$, $\lambda_9$, $\lambda_{10}$ , $\lambda_{13}$, $\lambda_{16}$) of the reduced density matrix $\rho$ for the non-contiguous bipartition of the periodic AKLT state as expressed in Eq. (\ref{eqn:lambda_noncontiguous1}). In panel (a) we consider the size of each $A$ blocks $\ell_A=2$. For moderately large $\ell_B$ these $\lambda_i$'s saturate to three non-unique values, given by Eq. (\ref{eqn:lambda_non_contiguous2}). In panel (b) we present the same plot but with a moderately large value of $\ell_A=20$. In this limit, for large $\ell_B$ all the $\lambda_i$'s converge to $\frac{1}{16}.$}
\label{fig:lambda_noncontiguous}
\end{figure} 

We start with the case when the sub-blocks of each partition are of equal size, i.e., $ l_{A_1}=l_{A_2}=\ell_A, l_{B_1}=l_{B_2}=\ell_B$. The entanglement spectrum of the reduced density matrix obtained in this case is given by (see Appendix \ref{AppendixA} for detailed derivations)
\begin{eqnarray}
\lambda_i&=&\frac{1}{16}(1-\gamma^{\ell_A})^2 (1-\gamma^{\ell_B})^2, i \in 1, 2, 3, 4, 5,\nonumber\\
 \lambda_j&=&\frac{1}{16}(1-\gamma^{\ell_A})^2  (1+3\gamma^{\ell_B}) (1-\gamma^{\ell_B}), j \in 6, 7 , 8, \nonumber\\
 \lambda_9&=&\frac{1}{2}(r-\sqrt{r^2-4s}),\nonumber\\
  \lambda_k&=&\frac{1}{16}(1+3\gamma^{\ell_A})(1-\gamma^{\ell_A}) (1-\gamma^{\ell_B})^2, k\in 10, 11, 12, \nonumber\\
\lambda_l&=& \frac{1}{16}(1+3\gamma^{\ell_A})(1-\gamma^{\ell_A}) (1+3\gamma^{\ell_B}) (1-\gamma^{\ell_B}), \nonumber\\&&
l\in 13, 14 , 15, \nonumber\\
\lambda_{16}&=&\frac{1}{2}(r+\sqrt{r^2-4s}), 
\label{eqn:lambda_noncontiguous1}
\end{eqnarray}
where
\begin{eqnarray}
s&=&\Big[\frac{1}{16}\Big((1+3\gamma^{\ell_A}) (1-\gamma^{\ell_A})(1+3\gamma^{\ell_B}) (1-\gamma^{\ell_B}) \Big)\Big]^2, \nonumber\\
r&=&\frac{1}{64} \Big((1+3\gamma^{\ell_A})^2(1+3\gamma^{\ell_B})^2+3(1+3\gamma^{\ell_A})^2(1-\gamma^{\ell_B})^2\nonumber\\
&+&3(1-\gamma^{\ell_A})^2(1+3\gamma^{\ell_B})^2+(1-\gamma^{\ell_A})^2(1-\gamma^{\ell_B})^2\Big).\nonumber\\
\end{eqnarray}
%
Hence, the entanglement spectrum in this case has six distinct eigenvalues. This is consistent with the expected degeneracy of the entanglement spectrum based on the SO(3) symmetry: $\frac{1}{2}\otimes \frac{1}{2} \otimes \frac{1}{2}\otimes \frac{1}{2}= 0 (2) \oplus 1 (3) \oplus 2(1)$. Among those $\lambda_{i}$'s, $\lambda_{16}$ is the highest one for all $\ell_B$. However, the lowest eigenstate depends on the parity of $\ell_B$. For even $\ell_B$, the degenerate set $\lambda_i$, with $i\in \{1,\cdots,5\}$ is the lowest one. Yet, for odd $\ell_B$, $\lambda_9$ becomes the lowest eigenvalue. Moreover, for a low $\ell_A$ and moderately large values of $\ell_B$ the number of such distinct eigenvalues decreases further and reduces to

\begin{eqnarray}
\lambda_i&\simeq& \frac{1}{16}(1-\gamma^{\ell_A})^2, i \in 1 ~\text{to}~9,\nonumber\\
\lambda_j&\simeq&\frac{1}{16}(1+3\gamma^{\ell_A})(1-\gamma^{\ell_A}), j\in 10 ~\text{to}~15, \nonumber\\
\lambda_{16}&\simeq&\frac{1}{16}(1+3\gamma^{\ell_A})^2. 
\label{eqn:lambda_non_contiguous2}
\end{eqnarray}
We observe that these $\lambda_i$ can be identified as products of the eigenvalues derived for the contiguous case, given in Eq. (\ref{eqn:lambda_contiguous}). This suggests that  in this limit,  the mixed state describing  part $A$ becomes approximately factorized, $\rho \approx \rho_{A_1}\otimes \rho_{A_2}$.
Therefore, the reduced density matrix $\rho$  becomes diagonal in the $\{|\eta_k^{\ell_A}\rangle\otimes |\eta_p^{\ell_A}\rangle\}$ basis, with $|\eta_k^{\ell_A}\rangle, |\eta_p^{\ell_A}\rangle\in \{|\phi^{\ell_A}_+\rangle, |\phi^{\ell_A}_-\rangle, |\phi^{\ell_A}_{01}\rangle, |\phi^{\ell_A}_{10}\}\rangle$. The eigenstate with highest eigenvalue, i.e., $\lambda_{16}$ can now be written as  $|\lambda_{16}\rangle=|\phi^{\ell_A}_+\rangle \otimes |\phi^{\ell_A}_+\rangle$  Furthermore, if we now increase the size of subsystem $A$ to a moderately high value, all the $\lambda_i$ approach $\frac{1}{16} (\epsilon_0=4\log2)$. We present the behavior of the $\lambda_i$ with parameters $\ell_A$ and $\ell_B$ in Fig. \ref{fig:lambda_noncontiguous}.

We are now equipped with the necessary tools to construct the entanglement Hamiltonian in this case. Similarly to the contiguous case, we expand the entanglement Hamiltonian $\mathcal{H}_{E}$ in the eigenbasis of the reduced density matrix obtained above and propose

\begin{eqnarray}
\mathcal{H}_E&=&-\ln(\rho),\nonumber\\
&=&\epsilon_0 + \Big(J_E^{(1)}\vec{\sigma}_1.\vec{\sigma}_2+J_E^{(2)}\vec{\sigma}_2.\vec{\sigma}_3+J_E^{(1)}\vec{\sigma}_3. \vec{\sigma}_4 +J_E^{(2)}\vec{\sigma}_4.\vec{\sigma}_1 \Big),\nonumber\\
\label{eqn:EH_non_contiguous}
\end{eqnarray}
where in the above expansion we only consider terms up to   two-body interactions with nearest-neighbor  exchange and  all other interaction terms have been dropped considering those as leading order approximation. Now for large values of $\ell_A$ and $\ell_B$, we have $\epsilon_0\approx  4 \log 2+O(\gamma^{\ell_A})$, $J_E^{(1)}\approx \gamma^{\ell_A}$, and $J_E^{(2)}\approx \gamma^{\frac{\ell_A+\ell_B}{2}}$.
Hence, in this case, the second part of the entanglement Hamiltonian can be understood as an interacting Hamiltonian, $H_{int}=\Big(J_E^{(1)}\vec{\sigma}_1.\vec{\sigma}_2+J_E^{(2)}\vec{\sigma}_2.\vec{\sigma}_3+J_E^{(1)}\vec{\sigma}_3. \vec{\sigma}_4+J_E^{(2)}\vec{\sigma}_4.\vec{\sigma}_1 \Big)$, where the effective spins at the end of each block interact with coupling strength $J_E^{(1)}$ and the inter-block interactions is given by $J_E^{(2)}$.  The illustration in Fig. \ref{fig:schematic_first} depicts the relation between different bipartitions of the AKLT state and the configuration of the interacting Hamiltonian derived from those. In this regard, we consider two limiting cases for $\ell_B$ and analyze the properties of the corresponding entanglement Hamiltonians in detail as follows.

\subsubsection{Large interblock spacing}

The first limit we consider is $\ell_B\gg \ell_A$, which results in $J_E^{(2)} \rightarrow 0$. Thus, the form of the GS of the $\mathcal{H}_E$ is given by
\begin{eqnarray}|\psi\rangle_G&=&|\phi^{\ell_A}_+\rangle \otimes |\phi^{\ell_A}_+\rangle, \nonumber\\&=&\left( \frac{|\phi^{\ell_A}_{00}\rangle+|\phi^{\ell_A}_{11}\rangle}{\sqrt{2}})\otimes (\frac{|\phi^{\ell_A}_{00}\rangle+|\phi^{\ell_A}_{11}\rangle}{\sqrt{2}}\right).
\end{eqnarray}
Now we carry out a similar basis transformation as the one in Eq. (\ref{eqn:Basis_transformation}) in Appendix\ref{AppendixA}, and write
\begin{eqnarray}|\psi\rangle_G&=&|\tilde{\phi}^{\ell_A}_+\rangle\otimes |\tilde{\phi}^{\ell_A}_+\rangle, \nonumber\\&=&\left( \frac{|\tilde{\phi}^{\ell_A}_{0 1}\rangle-|\tilde{\phi}^{\ell_A}_{10}\rangle}{\sqrt{2}})\otimes(\frac{|\tilde{\phi}^{\ell_A}_{01}\rangle-|\tilde{\phi}^{\ell_A}_{10}\rangle}{\sqrt{2}}\right).
\label{ground_state}
\end{eqnarray}
This provides a {\em spin-dimer}-like interpretation of the GS of the entanglement Hamiltonian $\mathcal{H}_E$. Therefore, the characteristic feature of the entanglement Hamiltonian remains consistent with the previous case. As we increase the number of boundaries of the subsystem of the AKLT state, the number of dimers in the GS of the entanglement Hamiltonian also increases proportionally.

The next excited state can be formed by breaking the dimer between any pair of sites (six such configurations, $\{|\tilde{\phi}^{\ell_A}_+\rangle \otimes |\eta_k^{\ell_A}\rangle, |\eta_k^{\ell_A}\rangle \otimes |\tilde{\phi}^{\ell_A}_+\rangle\}$, with $|\eta_k^{\ell_A}\rangle \in \{|\tilde{\phi}^{\ell_A}_-\rangle, |\tilde{\phi}^{\ell_A}_{01}\rangle, |\tilde{\phi}^{\ell_A}_{10}\}\rangle$) and finally the highest excited states are given by breaking both dimers (nine such configurations, $\{|\eta_k^{\ell_A}\rangle \otimes |\eta_p^{\ell_A}\rangle\}$, with $|\eta_k^{\ell_A}\rangle, |\eta_p^{\ell_A}\rangle \in\{|\tilde{\phi}^{{\ell_A}}_-\rangle, |\tilde{\phi}^{{\ell_A}}_{01}\rangle, |\tilde{\phi}^{\ell_A}_{10}\rangle\}$).

\subsubsection{Block size equals interblock spacing}

Let us now consider the case in which all the blocks have equal size, $\ell_A=\ell_B=l$, for even $l$, so the coefficients $J_E^{(1)}$ and $J_E^{(2)}$ become antiferromagnetic. In this case, the corresponding entanglement Hamiltonian  $\mathcal{H}_E$ becomes transnationally invariant and its GS matches exactly the GS of the physical Heisenberg Hamiltonian of four sites. Hence, we argue that the entanglement Hamiltonian derived from the equal-size non-contiguous bipartitions of the AKLT state eventually becomes an antiferromagnetic Heisenberg Hamiltonian in the auxiliary basis, which is a critical model. We elaborate on the analysis as follows.  

\begin{figure}[h]
\includegraphics[width=7.cm]{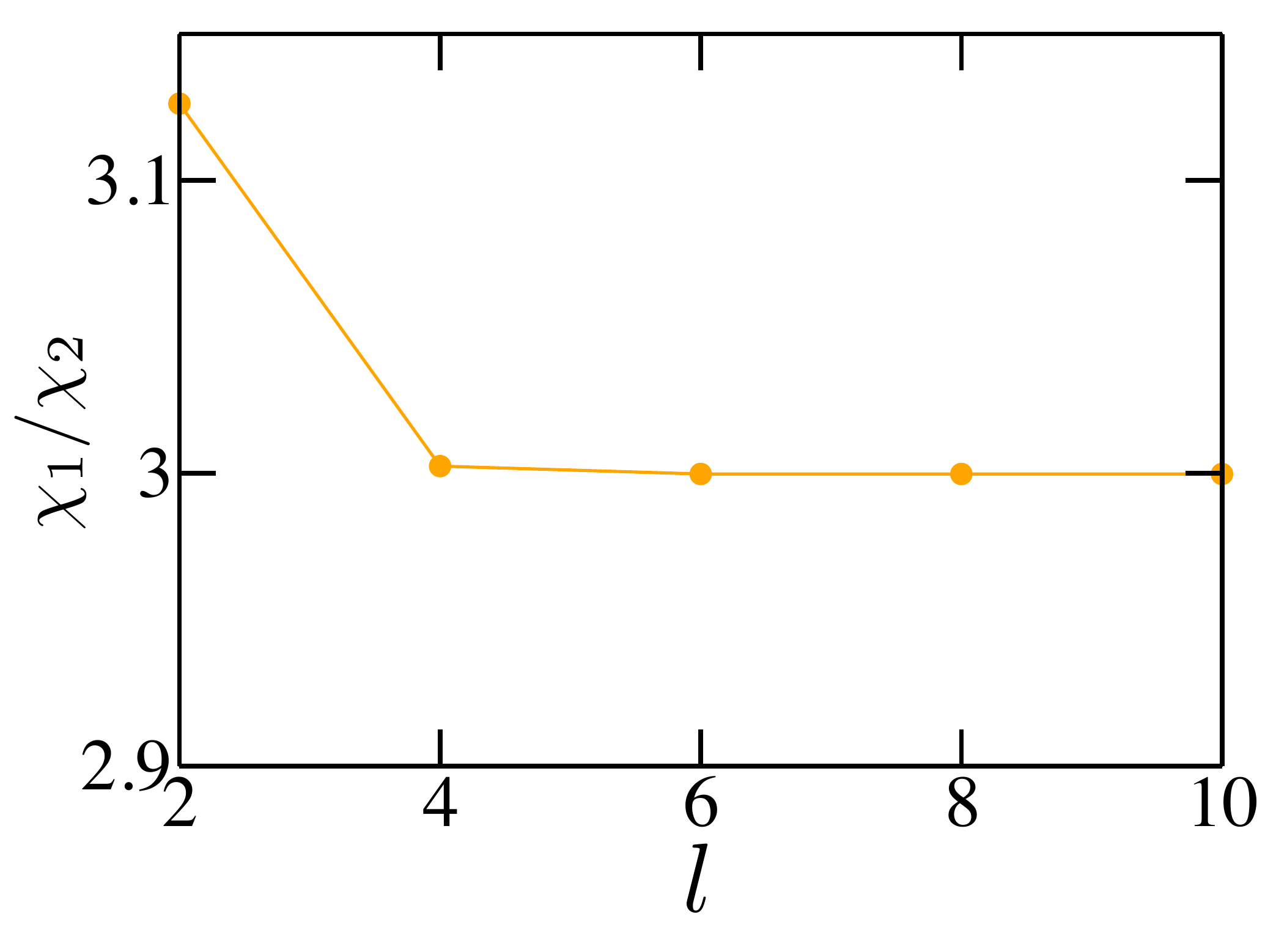}
\caption{Ratio between the coefficients $\chi_1$ and $\chi_2$ as expressed in Eq. (\ref{eqn:GS:Heisenberg}), as a function of the size of the $A$ and $B$ blocks $\ell_A=\ell_B=l$.}
\label{fig:coefficent_ratio}
\end{figure}

As a first step, we find that for $\ell_A=\ell_B=l$ the GS of the entanglement Hamiltonian given in Eq. (\ref{eqn:EH_non_contiguous}) can be expressed in the basis $\{|\phi^{l}_+\rangle, |\phi^{l}_-\rangle, |\phi^{l}_{01}\rangle, |\phi^{l}_{10}\rangle$ as
\begin{eqnarray}
|\psi\rangle_{G}&=& \chi_1 |\phi^{l}_+\rangle |\phi^{l}_{+}\rangle +\chi_2( |\phi^{l}_{-}\rangle |\phi^{l}_{-}\rangle+ |\phi^{l}_{01}\rangle |\phi^{l}_{10}\rangle\nonumber\\&+& |\phi^{l}_{10}\rangle |\phi^{l}_{01}\rangle),
\label{eqn:GS:Heisenberg}
\end{eqnarray}
where the ratio $\chi_1/\chi_2\rightarrow 3$ as we increase the size of the blocks, $l$. The behavior of the ratio $\chi_1/\chi_2$ with $l$ is depicted in Fig. \ref{fig:coefficent_ratio}. Now using $|\phi^{l}_{\pm}\rangle=\frac{1}{\sqrt{2}} (|\phi^{l}_{00}\rangle\pm|\phi^{l}_{11}\rangle)$ we can further decompose $|\psi\rangle_{G}$, obtaining
\begin{eqnarray}
|\psi\rangle_{G}&=&\frac{\chi_1 }{2}( |\phi^{l}_{00}\rangle+|\phi^{l}_{11}\rangle) ( |\phi^{l}_{00}\rangle+|\phi^{l}_{11}\rangle)\nonumber\\&+& \frac{\chi_2}{2}( |\phi^{l}_{00}\rangle-|\phi^{l}_{11}\rangle) ( |\phi^{l}_{00}\rangle-|\phi^{l}_{11}\rangle)\nonumber\\ &+&\chi_2(|\phi^{l}_{01}\rangle |\phi^{l}_{10}\rangle+ |\phi\rangle^{l}_{10} |\phi\rangle^{l}_{01}),\nonumber\\
&=& \frac{(\chi_1+\chi_2)}{2}(|\phi^{l}_{00}\rangle |\phi^{l}_{00}\rangle + |\phi^{l}_{11}\rangle |\phi^{l}_{11}\rangle)\nonumber\\&+&\frac{(\chi_1-\chi_2)}{2} (|\phi^{l}_{00}\rangle |\phi^{l}_{11}\rangle + |\phi^{l}_{11}\rangle |\phi^{l}_{00}\rangle)\nonumber\\&+&\chi_2(|\phi^{l}_{01}\rangle |\phi^{l}_{10}\rangle+ |\phi^{l}_{10}\rangle |\phi^{l}_{01}\rangle).
\end{eqnarray}
Using $\chi_1/\chi_2=3$, we get
\begin{eqnarray}
|\psi\rangle_{G}&=& 2\chi_2 (|\phi^{l}_{00}\rangle |\phi^{l}_{00}\rangle + |\phi^{l}_{11}\rangle |\phi^{l}_{11}\rangle)+\chi_2 (|\phi^{l}_{00}\rangle |\phi^{l}_{11}\rangle \nonumber\\&+&|\phi^{l}_{11}\rangle |\phi^{l}_{00}\rangle)+\chi_2 (|\phi^{l}_{01}\rangle |\phi^{l}_{10}\rangle+ |\phi^{l}_{10}\rangle |\phi^{l}_{01}\rangle).
\end{eqnarray}

Let us perform the basis transformation given in Eq. (\ref{eqn:Basis_transformation}), and rewrite the state $|\psi\rangle_{G}$ in the transformed basis as
\begin{eqnarray}
|\psi\rangle_{G}&=& 2\chi_2 \Big(|\tilde{\phi}^{l}_{01}\rangle |\tilde{\phi}^{l}_{01}\rangle + |\tilde{\phi}\rangle^{l}_{10}|\tilde{\phi}^{l}_{10}\rangle\Big)\nonumber\\&-&\chi_2\Big(|\tilde{\phi}^{l}_{01}\rangle |\tilde{\phi}^{l}_{10}\rangle +|\tilde{\phi}^{l}_{10}\rangle |\tilde{\phi}^{l}_{01}\rangle\Big)\nonumber\\&-&\chi_2\Big(|\tilde{\phi}^{l}_{00} \rangle|\tilde{\phi}^{l}_{11}\rangle+ |\tilde{\phi}^{l}_{11}\rangle  |\tilde{\phi}^{l}_{00}\rangle\Big).
\end{eqnarray} 
Now from normalization, we get $\chi_2=\frac{1}{2\sqrt{3}}$, which finally yields

\begin{eqnarray}
|\psi\rangle_{G}&=& \frac{1}{\sqrt{3}} \Big(|\tilde{\phi}^{l}_{01}\rangle |\tilde{\phi}^{l}_{01}\rangle + |\tilde{\phi}\rangle^{l}_{10}|\tilde{\phi}^{l}_{10}\rangle\Big)\nonumber\\&-&\frac{1}{2\sqrt{3}} \Big(|\tilde{\phi}^{l}_{01}\rangle |\tilde{\phi}^{l}_{10}\rangle +|\tilde{\phi}^{l}_{10}\rangle |\tilde{\phi}^{l}_{01}\rangle\Big)\nonumber\\&-&\frac{1}{2\sqrt{3}} \Big(|\tilde{\phi}^{l}_{00} \rangle|\tilde{\phi}^{l}_{11}\rangle+ |\tilde{\phi}^{l}_{11}\rangle  |\tilde{\phi}^{l}_{00}\rangle\Big).
\end{eqnarray} 
The above expression for $|\psi\rangle_G$ exactly matches the GS of a four site periodic spin-1/2 Heisenberg Hamiltonian, which in computational basis reads as

\begin{eqnarray}
|\psi\rangle_G^{Heisn}&=& \frac{1}{\sqrt{3}} \Big(|0101\rangle +|1010\rangle\Big)-\frac{1}{2\sqrt{3}}\Big(|0110\rangle+|1001\rangle\Big)\nonumber\\&-&\frac{1}{2\sqrt{3}} \Big(|0011\rangle+|1100\rangle\Big).
\end{eqnarray}
Therefore, from the above analysis, we conjecture that for any spin-1 quantum many-body state with an SPT order the entanglement Hamiltonian derived from its   non-contiguous partitions of equal size and with even number of sites is a  critical spin-1/2  antiferromagnetic Heisenberg Hamiltonian. If the model has $SU(2)$ symmetry  the entanglement Hamiltonian will belong to the 
universality class described by the WZW model $SU(2)_1$. SPT phases  in 1 + 1 dimensions have been analyzed using boundary CFT (BCFT) in references \cite{ent_ham_conformal3} and  \cite{SPT_CFT}.  In this approach the entanglement spectrum corresponds to a bipartition of the system, while in our case the connection between SPT phases and CFT arises from non-contiguous ones.  One should expect  a relationship between these two approaches.

Note that one can similarly consider another limit of the parameters $l_{A_1}, l_{A_2}, l_{B_1}, l_{B_2}$, such that $l_{A_1}=l_{A_2}=\ell_A$, $l_{B_1}=l_B$ and $l_{B_2} \rightarrow \infty$. In this limit, all the characteristic features remain as  above. The eigenspectrum for the $A_1A_2: B_1B_2$ bipartition is derived in Appendix \ref{AppendixA}. 

\section{Bulk-edge correspondence}
\label{Sec3}

The characterization of the entanglement Hamiltonians in the previous section hints towards its underlying connection to the physical Hamiltonian of the system, which can be unveiled through the bulk-edge correspondence that we will discuss in detail in this section. In order to unveil the bulk-edge correspondence in our case, we first consider the entanglement Hamiltonian derived from a contiguous bipartition of the periodic AKLT state and compare its low-energy behavior to that of the AKLT model with open boundary conditions, which we call the {\em physical Hamiltonian}. For an open AKLT model, the GS is four-fold degenerate consisting of a spin-triplet and a singlet. This GS degeneracy is exact for all system sizes. 
\begin{figure}[h]
\includegraphics[width=4cm]{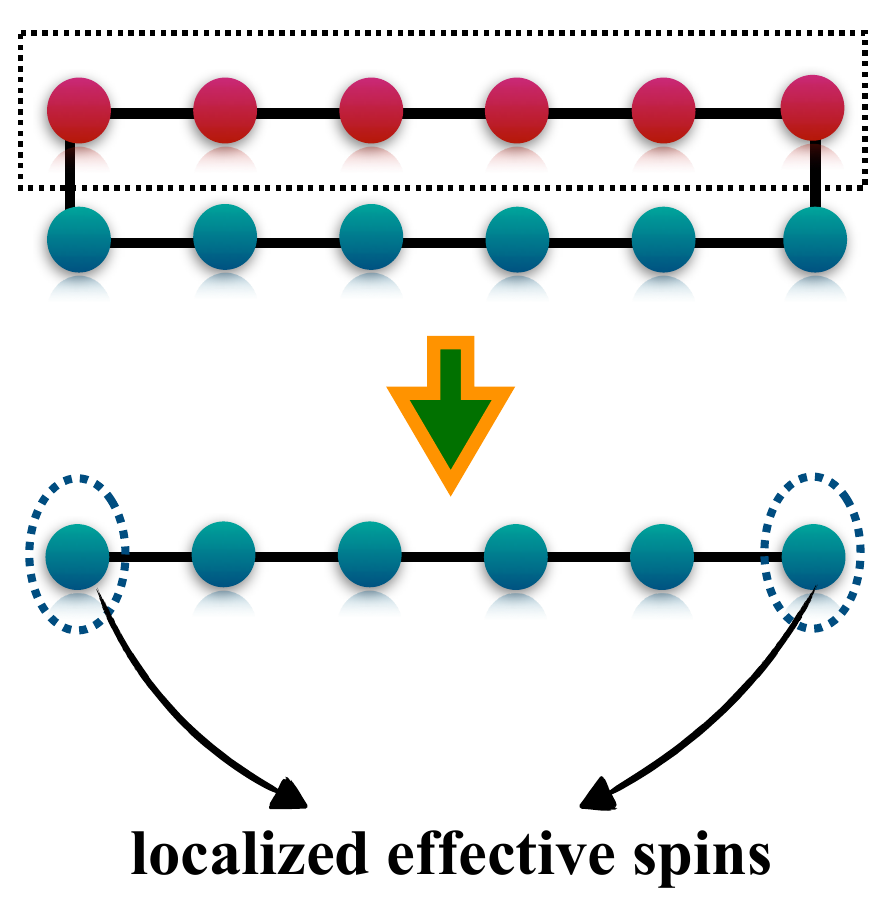}
\caption{Schematic of the edge degrees of freedom that appear at the end of the reduced block once we perform  the tracing operation in a periodic chain.}
\label{fig:schematic_second}
\end{figure} 
The GS configuration, in this case, can be well-explained in terms of the edge-degrees of freedom appearing at the end of the open chain \cite{AKLT2}. Loosely speaking, for a finite length open AKLT chain, the spin-1/2 degrees of freedoms at the edges combine to form a singlet and a triplet configuration, yielding a unique GS in the thermodynamic limit. 
Interestingly, from the derivations of the entanglement Hamiltonian as presented in Sec. \ref{Contiguous bipartition}, we observe that the low-energy part of the entanglement Hamiltonian coincides with that of the physical Hamiltonian. The low-energy subspace, in this case, is also spanned by a singlet and a triplet formed between the auxiliary spin-1/2 particles at the boundaries of the blocks (see Fig. \ref{fig:schematic_second} for a schematic representation). This justifies the emergence of the bulk-edge correspondence in this scenario. However, unlike the physical Hamiltonian, in this case, the degeneracy is not exact and the entanglement gap ($\Delta_{ent}$) closes only at moderately large value of   subsystem length $l$  (from Eq. (\ref{eqn:eigen_ent_Ham}) we can see that $\Delta_{ent}$  scales as $\Delta_{ent}\propto\gamma^l$).

We next move one step further and check whether the bulk-edge correspondence persists as we move away from the AKLT point. Towards this aim, we consider the bilinear-biquadratic Heisenberg (BBH) Hamiltonian, expressed in Eq. (\ref{eqn:BBH}). In our work, we mainly focus our study in the region $-\frac{\pi}{4}<\theta<\frac{\pi}{4}$, i.e., the Haldane phase of the model  \cite{SOP}. 
The physical Hamiltonian, in this case, is the open BBH Hamiltonian for the region $-\frac{\pi}{4}<\theta<\theta_{AKLT}$, whose GS is always either a singlet for an even number of sites, and a triplet otherwise \cite{AKLT2}. In our case, without loss of generality, we always choose the size of the physical Hamiltonian ($N_O$)  to be even.

\begin{figure}[h]
\includegraphics[width=7cm]{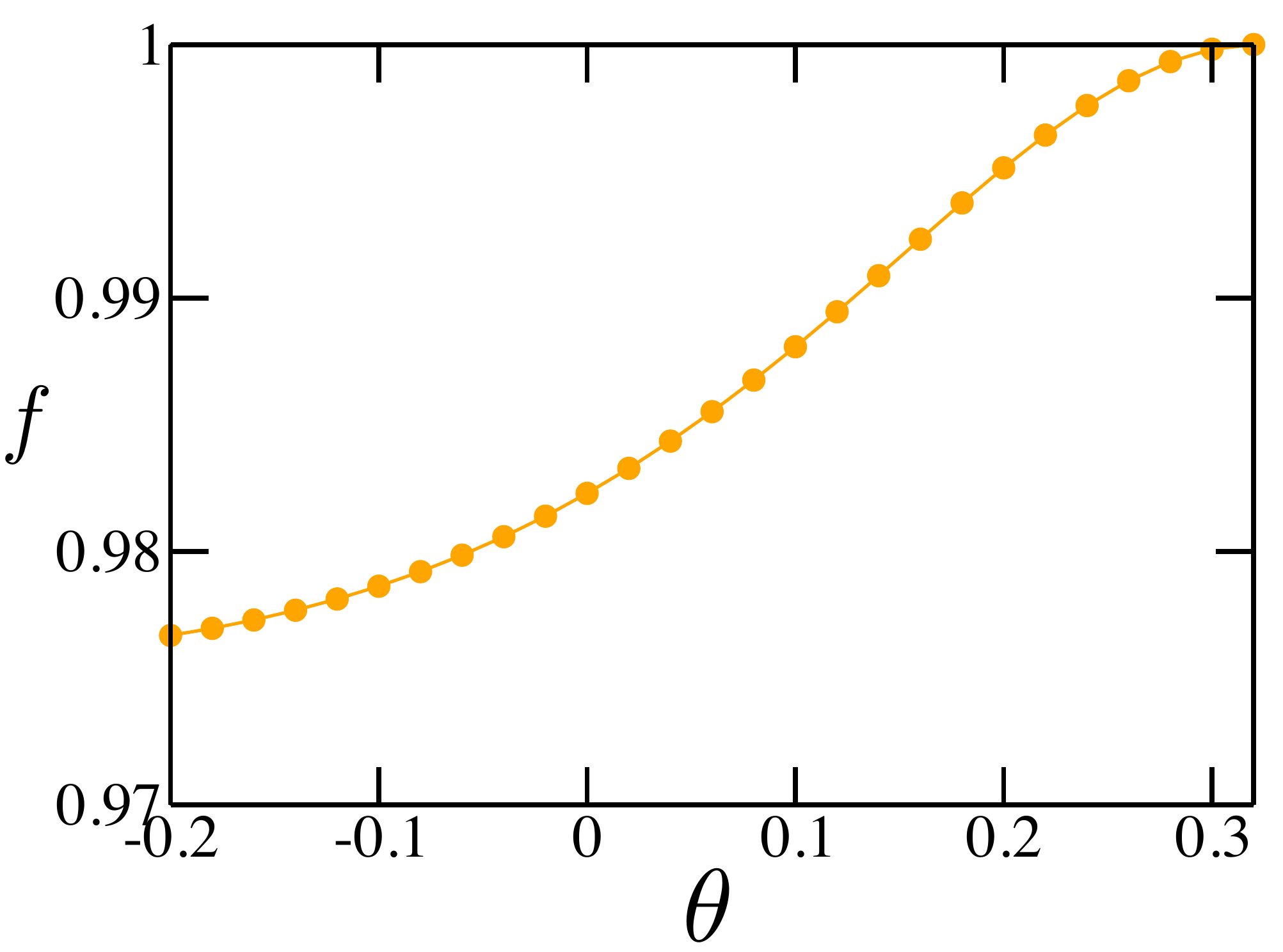}
\caption{Fidelity between the GS of the physical open BBH Hamiltonian and the GS of the entanglement Hamiltonian for the contiguous bipartition of the ground state of the periodic BBH Hamiltonian. Here,  the system size for the physical Hamiltonian is $N_O=4$ and entanglement Hamiltonian of size  $l=4$ is  obtained from a periodic chain with $N_P=8$.} 
\label{fig:fidelity}
\end{figure} 

To establish the bulk-edge correspondence at other points in the Haldane phase, we compute the reduced density matrix of the GS of the periodic BBH Hamiltonian in the region $-\frac{\pi}{4}<\theta<\theta_{AKLT}$.   We denote the size of the PBC Hamiltonian by $N_P$. Unfortunately, an analytical derivation of the reduced density matrix is not available away from the AKLT point. As a result, we first compute the ground state of the model by employing the density matrix renormalization group (DMRG) \cite{DMRG1,DMRG2}.  Here we use the finite-size algorithm with an adaptable number of kept states and a fixed tolerance on the discarded weights of the density matrix.  The entanglement Hamiltonian ($\mathcal{H}_E$) is then constructed from the reduced density matrix obtained from the contiguous partition of the ground state and its eigenenergies are derived from the eigenvalues of the reduced density matrix $\lambda_i$, $e^{ent}_i=-\log(\lambda_{m-i})$, where $i\in 0,1,2,\dots, m, m\approx 100$  and the $\lambda_i$ are arranged in ascending order. The low-energy part of the entanglement Hamiltonian is then analyzed and compared to that of the physical Hamiltonian of the same size ($N_O=l=N_P/2$).  We now discuss the results of such comparison in detail.

We find that the GS of the entanglement Hamiltonian for even $l$ is a singlet in
the considered Haldane region ($-\frac{\pi}{4}<\theta<\theta_{AKLT}$),
similarly to the physical Hamiltonian. The fidelity between the GS of the physical Hamiltonian $|\psi_0^{phy}\rangle$ and the entanglement Hamiltonian $|\psi_0^{ent}\rangle$, $f=|\langle \psi^{phy}_0|\psi_0^{ent}\rangle|^2$ in the region of interest is depicted in Fig. \ref{fig:fidelity}. From the plot we can see that the fidelity remains high throughout the considered region and decreases monotonously as $\theta$ moves away from the AKLT point. This provides a strong justification of the proposed bulk-edge correspondence: the high overlap of both GS points at the fact that the low-energy part of both Hamiltonians is structurally equivalent.\\

\begin{figure}[h]
\includegraphics[width=8.5cm]{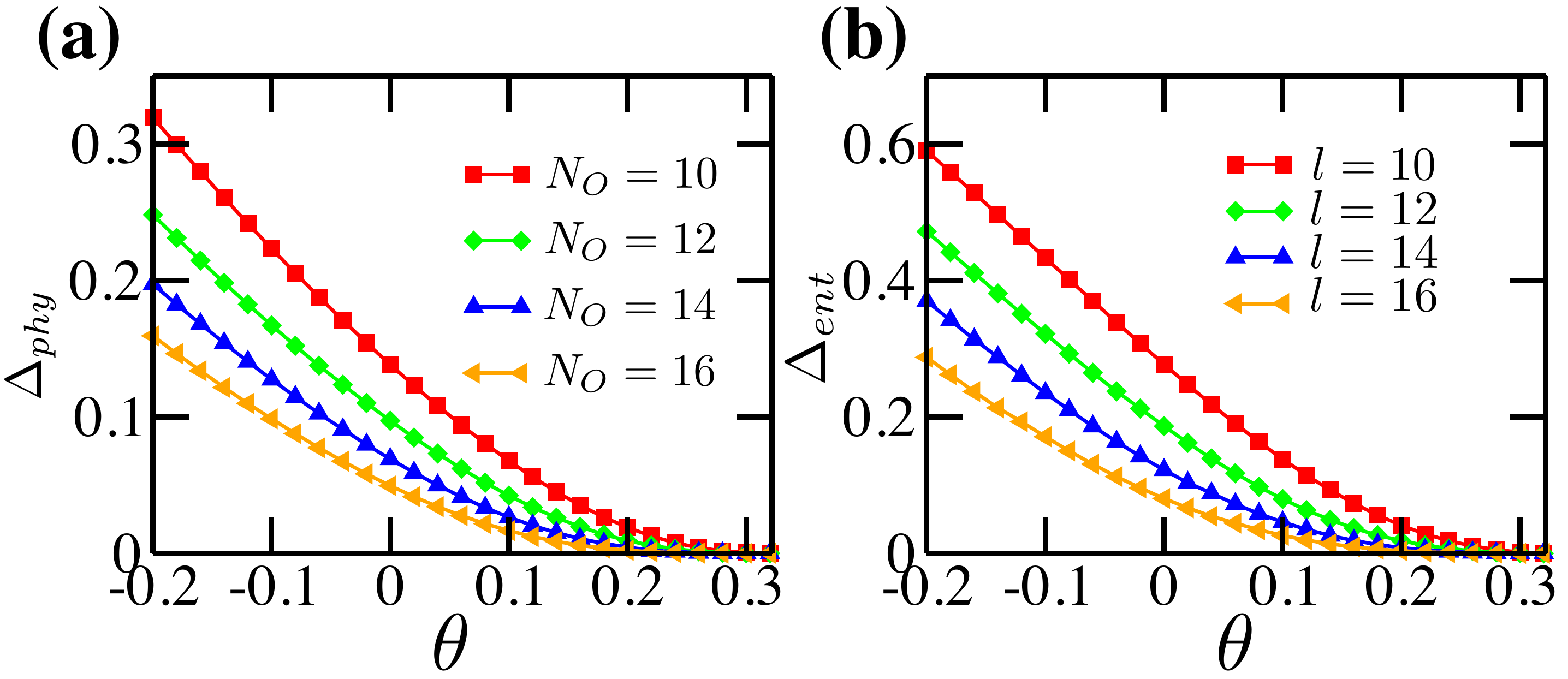}
\caption{(a)  Physical gap ($\Delta_{phy}$) and (b) entanglement gap ($\Delta_{ent}$) as a function of $\theta$ for various system sizes $N_O=l=N_P/2=10,12,14,16$. To compute $\Delta_{ent}$ we consider the length of the periodic BBH Hamiltonian ($N_P$) is double to that of the physical Hamiltonian ($N_O$).}
\label{fig:phy_gap_ent_gap}
\end{figure} 

We next establish the bulk-edge correspondence through the identical behavior of the energy gaps obtained for both Hamiltonians. Away from the AKLT point the exact degeneracy of the GS of the physical Hamiltonian $H_{BBH}$ breaks, giving rise to an energy gap which decays exponentially with the system size, $\Delta_{phy} =a_0(\theta,N_O) \exp\(-N_O/\xi_{phy}(\theta,N_O)\)$ \cite{AKLT2}, where a correlation length is introduced, $\xi_{phy}$, with an explicit dependence on $\theta$ and $N_O$. In case of the entanglement Hamiltonian, the gap is defined as $\Delta_{ent}=e^{ent}_1-e^{ent}_0=\log(\lambda_{m})-\log(\lambda_{m-1})$. 
We plot $\Delta_{ent}$ as a function of $\theta$ for different sizes $l$  and compare it with $\Delta_{phy}$ in Fig. \ref{fig:phy_gap_ent_gap}. One can note that both energy gaps remain very close to each other, hinting towards a similar decay factor. To confirm such behavior, we have performed a finite size scaling analysis of $\Delta_{ent}$, and observed that the entanglement gap scales as $\Delta_{ent}\propto \exp\(-l/\xi_{ent}(\theta, l)\)$.

\begin{figure}[t]
\includegraphics[width=8.5cm]{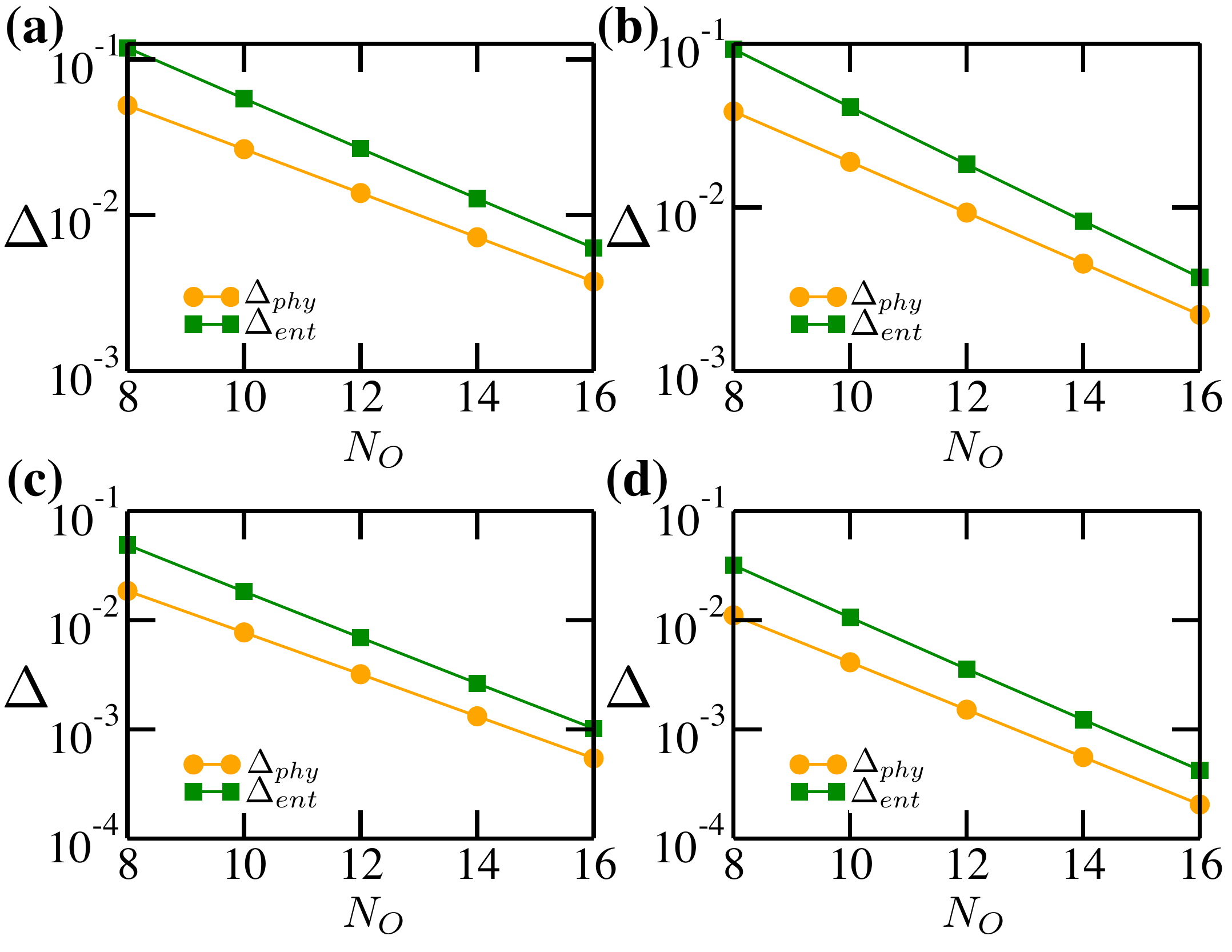}
\caption{Physical gap ($\Delta_{phy}$) and entanglement gap ($\Delta_{ent}$) as a function of the system size ($N_O=l=N_P/2$)    for different regions of the Haldane phase (a) $\theta=0.18$, (b) $\theta=0.20$, (c) $\theta=0.24$, and (d) $\theta=0.26$. Notice that the vertical axis is scaled logarithmically.}
\label{fig:phy_gap_ent_gap_scaling}
\end{figure}

The scaling behavior of the energy gaps of both Hamiltonians with the system size  for a few points in the Haldane phase is depicted in Fig.  \ref{fig:phy_gap_ent_gap_scaling}, where panels (a)-(d) are obtained for four different values of $\theta$, eg., (a) $\theta=0.18$, (b) $\theta=0.20$, (c) $\theta=0.24$, (d) $\theta=0.26$, and the $y$-axis in all the plots is in log-scale. We observe that as $\theta$ approaches the AKLT point, the plots become more parallel to each other. 
\begin{table}[h]
\centering
 \begin{tabular}{||c ||c|| c||} 
 \hline
 \thead{$\theta$} & \thead{$\xi_{phy}(\theta, \infty)$} & $\xi_{ent}(\theta, \infty)$\\ [0.5ex] 
 \hline\hline
 0.00 & 5.232&5.138 \\
   \hline \hline
0.06 & 4.549  &4.233 \\
  \hline \hline
0.10 & 4.070&  3.668\\
\hline \hline
 0.16&  3.323 & 2.908   \\
  \hline \hline
0.20 & 2.812  &2.445  \\
  \hline \hline
 0.24&2.293 & 2.006\\
  \hline \hline
   0.26&  2.028& 1.788 \\
   \hline \hline
 0.28&1.785 & 1.564 \\
 \hline \hline
 0.30& 1.511 & 1.320\\
 \hline \hline
 0.31& 1.369& 1.173\\
[1ex] 
 \hline
 \end{tabular}
\caption{Correlation lengths $\xi_{phy}(\theta,\infty)$ and $\xi_{ent}(\theta,\infty)$ for increasing values of $\theta$, obtained through the fitting of  the corresponding  energy gaps.}
  \label{Table1}
\end{table}
This can be confirmed from the behavior of the quantities $\xi_{phy}(\theta,\infty)$ and $\xi_{ent}(\theta,\infty)$, which correspond to the correlation lenghts in the thermodynamic limit, obtained from the  fitting of the plots, as shown in Table \ref{Table1}. From the table, one can see that $\xi_{phy}(\theta, \infty)$ and $\xi_{ent}(\theta,\infty)$ converge to  almost the same value, $\frac{1}{\log(3)}=0.9102$ as we approach the AKLT point \cite{AKLT2}.  Hence, at the AKLT point though the system has an exact degeneracy in the physical spectrum, the correlation length remains non-zero, $\xi_{phy}(\theta_{AKLT}, \infty)=0.9102$. It is the coefficient $a_0$ in $\Delta_{phy}$  that vanishes at that point.

We have found that the bulk-edge correspondence  also manifests itself in the properties of the low-energy excited states. For instance, we note that the degeneracies of the five lowest eigenenergies for the physical Hamiltonian ($e_0^{phy}, \dots, e_4^{phy}$) and the entanglement Hamiltonian ($e_0^{ent}, \dots,  e_4^{ent}$) coincide: $(1)$, $(3)$, $(5)$, $(3)$, $(3)$. This, in other words, certifies that  not only the ground states but the lowest  15 $(1+3+5+3+3=15)$ eigenstates of both the Hamiltonians  are also indeed qualitatively similar. Let us recall that the first two degeneracy factors, $(1)$ and $(3)$, correspond to the GS   singlet and triplet configurations respectively, which we have discussed above in detail. We may interpret the remaining multiplets ($(5)$, $(3)$ and $(3)$) of the physical Hamiltonian as follows. The periodic physical Hamiltonian has a triplet excitation above the singlet. Thus, one could expect that for an open physical chain there this triplet will combine with the edge spins. These edge spins will be combined as a singlet and a triplet: 1/2 $\otimes$ 1/2=0 $\oplus$ 1. Further combination with the excited triplet state finally yields  (0 $\oplus$ 1) $\otimes$ 1 = 0 $\oplus$ 1 $\oplus$  1 $\oplus$ 2. Hence, we can relate the degeneracy factors with their total spin as $(d_f, s)\to (5,2), (3,1), (3,1), (1,0)$. Note that the singlet $(1, 0)$ arising in the above decomposition remains just above these multiplets ($(5)$, $(3)$ and $(3)$) of physical spectrum  and does not appear in the  entanglement spectrum.  
\begin{figure}[h]
\includegraphics[width=8.5cm]{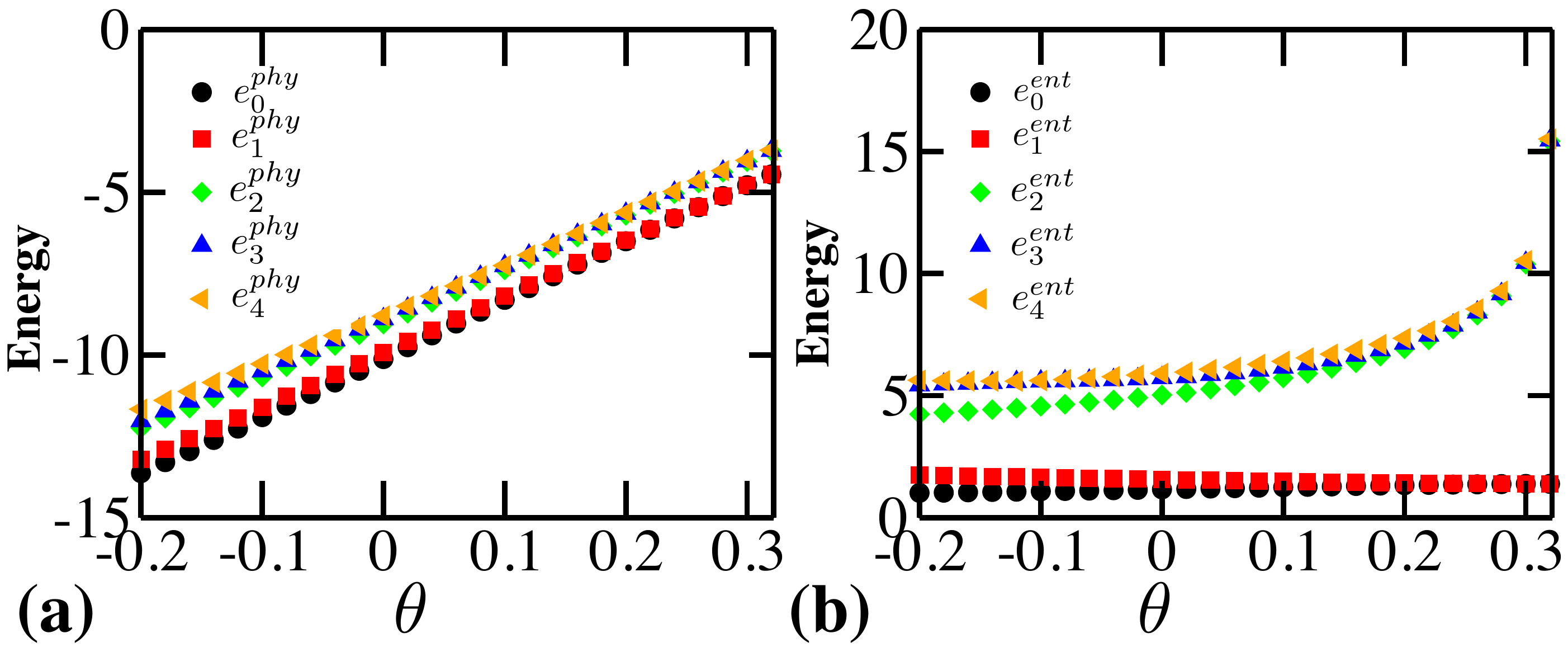}
\caption{The five lowest energy eigenvalues of (a) the physical Hamiltonian ($e_0^{phy}, \dots, e_4^{phy}$) and (b) the entanglement Hamiltonian ($e_0^{ent}, \dots,  e_4^{ent}$) as a function of $\theta$. This certifies  a strong correspondence exists even up to 15 lowest eigenstates of both the Hamiltonians. Here, we consider $N_O=l=N_P/2=8$.}
\label{fig:eign_spectrum}
\end{figure} 
Hence, starting from there, the degeneracy factors  of both  Hamiltonians differ.
It can be observed that not only the degeneracies coincide: the behavior of the set of eigenvalues as a function of $\theta$ remains qualitatively similar, as presented in Fig. \ref{fig:eign_spectrum}. Indeed, as we approach the AKLT point we observe how the eigenenergies of both Hamiltonians become degenerate following two different paths. Thus, the ground state ($e_0$) and first excited state ($e_1$) converge together whereas the next three excited energies ($e_2$, $e_3$ and $e_4$) follow a different route and finally converge as we approach the AKLT point. However, it is interesting to note that for the entanglement Hamiltonian, as we approach the AKLT point, the different paths are different from those of the entanglement Hamiltonian. This observation can be given a renormalization group (RG) interpretation, where the AKLT state represents the fixed point of an RG obtained by  keeping only the most relevant degrees of freedoms. As a result,  the angle $\theta$ in  Fig. \ref{fig:eign_spectrum} (b) can be identified with the RG direction since only the lowest eigenstates survives as $\theta \rightarrow \theta_{AKLT}$.

\begin{table}[h!]
\centering
\begin{adjustbox}{width=0.49\textwidth}
 \begin{tabular}{||c ||c|| c ||} 
 \hline
 \thead{Low energy \\ properties} & \thead{$\mathcal{H}_E(l)$} & \thead{$H_{BBH}(N_O)$}\\ [0.5ex] 
 \hline\hline
  \thead{Ground state configuration}  &\makecell{singlet (even $l$), \\ triplet (odd $l$)} &  \makecell{singlet (even $N_O$), \\ triplet (odd $N_O$)}\\
  \hline \hline
\thead{Scaling of energy gap}  & \makecell{$\Delta_{ent}\propto \exp\(-l/\xi_ {ent}\)$}& \makecell{ $\Delta_{phy}\propto \exp\(-N_O/\xi_{phy}\)$}  \\
  \hline \hline
   \thead{Correlation length} & $\xi_{ent}(\theta\rightarrow\theta_{AKLT},l)\rightarrow \frac{1}{\log(3)}$ &  $\xi_{phy}(\theta\rightarrow\theta_{AKLT},N_O)\rightarrow \frac{1}{\log(3)}$\\
\hline \hline
  \thead{Degeneracy factor (.)}  & (1), (3), (5), (3), (3) & (1), (3), (5), (3), (3)\\
  [1ex] 
 \hline
 \end{tabular}
 \end{adjustbox}
 \caption{ In this table, we summarize the correspondence between  the low-energy  properties of the  entanglement Hamiltonian ($\mathcal{H}_E$) and corresponding physical Hamiltonian $H_{BBH}$ of same length $N_O=l=N_P/2$.}
  \label{Table:BE_summary}
\end{table}

Therefore, our analysis supports the conjecture that a bulk-edge correspondence emerges between the entanglement Hamiltonian derived from a part of our quantum many-body state and its corresponding physical Hamiltonian.
We summarize all the key features of the proposed bulk-edge correspondence discussed above in Table \ref{Table:BE_summary}.

\section{String order parameter}
\label{sec:SOP}
\begin{figure}[t]
\includegraphics[width=7.5cm]{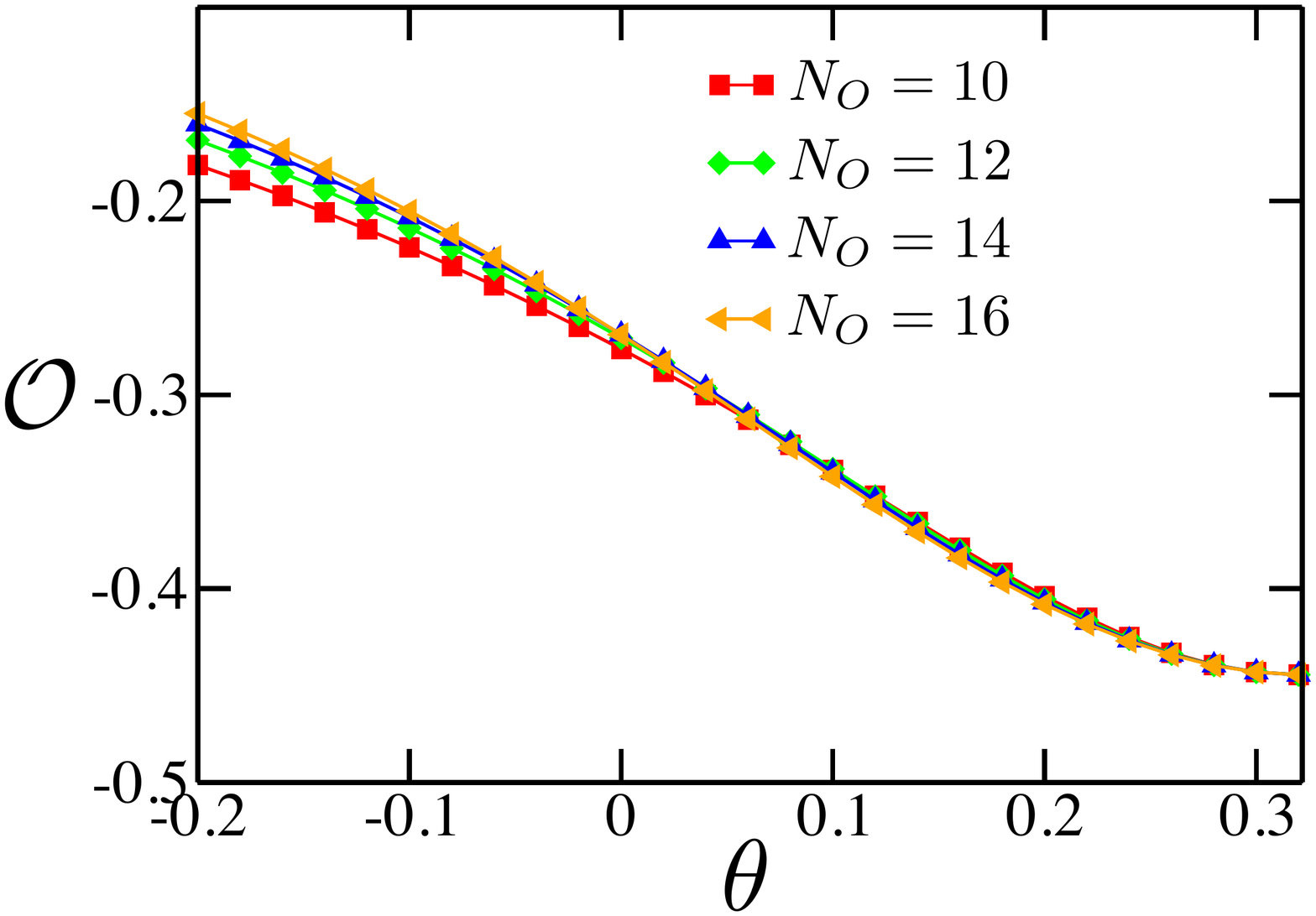}
\caption{String order parameter ($\mathcal{O}$) as defined in Eq. (\ref{eqn:SOP}) as a function of $\theta$ for different sizes of the open BBH chain,   $N_O =10$, $12$, $14$, $16$. In all the cases $\mathcal{O}$ have been computed for the spins situated at site ‘1’ and the end site  ‘$N_O$’. Hence, here ‘$l=N_O-2$’.} 
\label{figure:SOP}
\end{figure}

Let us compare the behavior of the energy gaps obtained above with other  relevant physical quantities computed for the model. In particular, we may consider the string order parameter (SOP) \cite{SOP} of the GS of the physical Hamiltonian, defined as  
\begin{eqnarray}
\mathcal{O}_{\theta}(l, N_O)=\langle  S^z_1 e^{-i\pi \sum_{k=2}^{l+1}  S^z_{k}} S^z_{l+2} \rangle.
\label{eqn:SOP}
\end{eqnarray}
The Haldane phase of the BBH Hamiltonian presents a hidden topological order which can not be detected using any conventional local order parameter. Yet, the non-local SOP takes non-zero values over the whole Haldane phase and acts as an order parameter distinguishing it from the other phases. We plot the behavior of SOP ($\mathcal{O}$) for our region of interest, $-\frac{\pi}{4}< \theta < \theta_{AKLT}$ and for different sizes of the open chain, in Fig. \ref{figure:SOP}. From the figure we note that the SOP ($\mathcal{O}$) attains its minimum value at the AKLT point and increases as it moves away from it, in similarity to the energy gaps. The finite-size scaling of the SOP presents some subtle points. We first consider its behavior at the AKLT point, for which an exact analytical form of $\mathcal{O}(N_O)$ can be obtained as follows.
Indeed, we may apply Eq. (\ref{eqn:SOP}) to any matrix product state using the tranfer matrix formalism, obtaining \cite{SOP_AKLT}  
\begin{eqnarray}
\mathcal{O}_{\theta}(l, N_O)= \text{Tr}(E^{N_O-l-2} \widehat{S^z} ({e^{-i \pi \widehat{S^z}} )^l\widehat{S^z}),} 
\label{eqn:SOP_MPS}
\end{eqnarray}

where $E$ is the transfer matrix defined earlier in Eq. (\ref{eqn:Transfer_matrix}), $\widehat{S^z}=\sum_{kk'} (A_k \otimes A^*_{k'}) \langle k|S^z|k'\rangle$, and $\tilde{E}={e^{-i \pi \widehat{S^z}}}=\sum_{kk'} (A_k \otimes A^*_{k'}) \langle k|e^{-i \pi S^z}|k'\rangle$. Notice that, in similarity to the transfer matrix, $\tilde{E}$ can be decomposed in its right and left eignevectors as follows, $\tilde{E}=\sum_{i=0}^{D^2-1} \tilde{\gamma}_s |\tilde{R}_s\rangle\langle\tilde{L}_s|$, with 

\begin{align}
|\tilde{R}_0\rangle=|R_1\rangle, \;\; \langle \tilde{L}_0|=\langle L_1|, \;\;
|\tilde{R}_1\rangle=|R_0\rangle, \;\; \langle \tilde{L}_1|=\langle L_0|,\nonumber\\
|\tilde{R}_2\rangle=|R_2\rangle, \;\; \langle \tilde{L}_2|=\langle L_2|, \;\;
|\tilde{R}_3\rangle=|R_3\rangle, \;\; \langle \tilde{L}_3|=\langle L_3|,\nonumber\\
\end{align}
and $\tilde{\gamma}_k=\gamma_k$, for all $k$, where the set of eigenvectors  and eigenvalues $\{|R_k \rangle, | L_k\rangle, \gamma_k\}$ obtained for the AKLT state was defined in Eq. (\ref{eqn:transfer_matrix_vec}). Using this, we can further simplify Eq. (\ref{eqn:SOP_MPS}) as follows.
\begin{eqnarray}
\mathcal{O}_{\theta_{AKLT}}(l, N_O) &=&  \text{Tr} \(\[ \gamma_k^{N_O-l-2}|R_k\rangle \langle L_k|\]\widehat{S^z} \[\sum_{s=0}^{D^2-1}\tilde{\gamma}^l_s |\tilde{R}_s\rangle\langle \tilde{L}_s| \]\widehat{S^z}\), \nonumber\\
&=&\sum_{k=0}^{D^2-1}\( \gamma_k^{N_O-l-2} \langle L_k|\widehat{S^z} \[\sum_{s=0}^{D^2-1}\tilde{\gamma}^l_s |\tilde{R}_s\rangle\langle \tilde{L}_s|\]\widehat{S^z}|R_k\rangle\),\nonumber\\
  &=&\langle L_0|\widehat{S_z}|\tilde{R}_0\rangle\langle \tilde{L}_0|\widehat{S^z}|R_0\rangle\nonumber\\
  &+&\sum_{s=1}^{D^2-1} \tilde{\gamma}_s^{l}\langle L_0|\widehat{S^z}|\tilde{R}_s\rangle\langle \tilde{L}_s|\widehat{S^z}|R_0\rangle\nonumber \\
  &+&\sum_{k=1}^{D^2-1}  \gamma^{N_O-l-2}_k\langle L_k|\widehat{S^z}|\tilde{R}_0\rangle\langle \tilde{L}_0|\widehat{S^z}|R_k\rangle\nonumber\\
  &+&\sum_{k,s=1}^{D^2-1} \gamma_k^{N_O-l-2} \tilde{\gamma}^l_s \langle L_k|\widehat{S^z}|\tilde{R}_s\rangle\langle \tilde{L}_s|\widehat{S^z}|R_k\rangle.
\end{eqnarray}
Now using the action  of  $\widehat{S^z}$ on $|R_i\rangle, |L_i\rangle$ and $|\tilde{R}_i\rangle, |\tilde{L}_i\rangle$ (see Appendix \ref{AppendixB}) we can further show that 
\begin{align}
\mathcal{O}_{\theta_{AKLT}}(l, N_O)=&-\(\frac{2}{3}\)^2 \( 1
  + \sum_{s=1}^{D^2-1} \tilde{\gamma}_s^{l} \delta_{s,0} 
  + \sum_{k=1}^{D^2-1}\gamma_k^{N_O-l-2} \delta_{k,0} \right.\nonumber\\
  &  \left. + \gamma_1^{N_O-l-2} \tilde{\gamma}^l_1 \), \nonumber\\
=&-\(\frac{2}{3}\)^2-\(\frac{2}{3}\)^2  \(-\frac{1}{3}\)^{N_O-2}.\nonumber\\
   \end{align}
  Now as $\mathcal{O}_{\theta_{AKLT}}(l, N_O)$ does not depend on $l$, we can replace  $\mathcal{O}_{\theta_{AKLT}}(l, N_O)$ by  $\mathcal{O}_{\theta_{AKLT}}(N_O)$   and get
   \begin{align}
\mathcal{O}_{\theta_{AKLT}}(N_O)=&-\(\frac{2}{3}\)^2-\(\frac{2}{3}\)^2  \(-\frac{1}{3}\)^{N_O-2}, \nonumber\\
=&-\(\frac{2}{3}\)^2-\(\frac{2}{3}\)^2 \(-\frac{1}{3}\)^{-2}\exp\(-\frac{N_O}{\frac{1}{\log(3)}}\), \nonumber\\
=&\mathcal{O}_{\theta_{AKLT}}(\infty)+\mathcal{A}(\theta_{AKLT},\infty) \exp\(-\frac{N_O}{\xi_{\mathcal{O}}(\theta_{AKLT}, \infty)}\), 
\label{eqn:SOP_fitting}
\end{align}
with $\mathcal{O}_{\theta_{AKLT}}(\infty)=-(\frac{2}{3})^2$, $\mathcal{A}(\theta_{AKLT},\infty)=-4$ and $\xi_{\mathcal{O}}(\theta_{AKLT}, \infty)=\frac{1}{\log(3)}=0.9102$. Hence, at the AKLT point, $\mathcal{O}_{\theta_{AKLT}}(N_O)$ approaches its thermodynamic limit $\mathcal{O}_{\theta_{AKLT}}(\infty)$ exponentially with the system size $N_O$. This behavior remains similar to the exponential decay of the energy gaps obtained above.  Interestingly, one can see that that the relation remains independent of the inter-spin distance $l$ and depends only on the total system size $N_O$.  We may conjecture is generic behavior of $\mathcal{O}_{\theta_{AKLT}}(l, N_O)$ still holds close to the $\theta_{AKLT}$, so Eq. (\ref{eqn:SOP_fitting}) can still be used to find $\xi_{\mathcal{O}}(\theta,\infty)$ for other points in the Haldane phase  lies in   the  neighborhood of  the AKLT point. Using this fact, we have obtained $\mathcal{O}_{\theta}(\infty)$, $\mathcal{A}(\theta,\infty)$ and $\xi_{\mathcal{O}}(\theta, \infty)$ for                    $\theta=0.28, 0.30, 0.31, 0.32$  in Table   \ref{Table3}. From the table, we can clearly see that as $\theta \rightarrow \theta_{AKLT}$, all the three-quantities  \big($\mathcal{O}_{\theta}(\infty)$, $\mathcal{A}(\theta,\infty)$, $\xi_{\mathcal{O}}(\theta, \infty)$\big) approach to their respective  values obtained at AKLT point.

\begin{table}[h]
\centering
 \begin{tabular}{||c ||c|| c||c||} 
 \hline
 \thead{$\theta$} & \thead{$\mathcal{O}_{\theta}(\infty)$} & $\mathcal{A}(\theta,\infty)$& $\xi_{\mathcal{O}}(\theta, \infty)$\\ [0.5ex] 
 \hline\hline
 0.28 & -0.438 &-1.27&1.106 \\
 \hline\hline
  0.30 & -0.443 &-1.39&1.062 \\
 \hline\hline
 0.31 & -0.444 &-1.57&1.027 \\
 \hline\hline
 0.32 & -0.444 &-2.19&0.944 \\
[1ex] 
 \hline
 \end{tabular}
 \caption{Behavior of the coefficients $\mathcal{O}_{\theta}(\infty)$, $\mathcal{A}(\theta,\infty)$ and $\xi_{\mathcal{O}}(\theta, \infty)$, obtained using  Eq. (\ref{eqn:SOP_fitting}) for $\theta$ values close to the AKLT point.}
 \label{Table3}
 \end{table}

 The scaling of both energy gaps and the SOP allows us to argue that the characteristic scale of the BBH model remains close to the value  $\frac{1}{\log(3)}$ and  manifests in the scaling of relevant physical quantities associated to it.

\section{Conclusions and Further Work}
\label{Sec4}

In this work, we have investigated the properties of the entanglement Hamiltonian for contiguous and non-contiguous bipartitions of the AKLT state. In both cases, the entanglement Hamiltonian can be expressed as an interacting Hamiltonian acting between the auxiliary spins at the edges of the blocks. In particular, when the block sizes are equal  and comprises of even number of sites  the corresponding entanglement Hamiltonian becomes the antiferromagnetic  Heisenberg model, which is critical.  We conjecture that this behavior is generic for other SPT states.

We further observed that the low-energy properties of the entanglement Hamiltonian obtained for the contiguous bipartition of the AKLT state can be related to the edge properties of the physical Hamiltonian with open boundaries, hinting towards a bulk-edge correspondence that has been extensively studied in many earlier works  \cite{ent_spectrum_ref5,ent_ham_ref2,ent_ham_ref3,ent_ham_ref4,ent_ham_ref5,ent_ham_ref6}. To investigate this bulk-edge correspondence in our context in more detail, we also analyzed the properties of the entanglement Hamiltonian obtained for other states within the Haldane phase. By employing the DMRG technique, we obtained the GS of the periodic BBH Hamiltonian and derived the corresponding entanglement Hamiltonian from a contiguous subsystem. The low-energy properties are then analyzed and compared to that of the physical model of the same size.   As a first result, we showed that for the whole region considered, the GS of the physical and entanglement Hamiltonians remain structurally equivalent, which serves as even a stronger example of the existing bulk-edge correspondence. Additionally, similarly to the physical gap, we found that the entanglement gaps decays exponentially with the system size, with the associated correlation lengths obtained from a finite-size scaling analysis approaching the same value for the AKLT point. Along with this, we showed that the bulk-edge correspondence manifests itself the properties of a few excited states of both Hamiltonians, as we can see in the identical degeneracies for the first five low-energy eigenstates, and their behavior in the region of the Haldane phase we have considered in our work. 

Finally, the behavior of the energy gaps is compared to that of the string order parameter. It is  shown that similar to the energy gaps, SOP decays exponentially towards its asymptotic as we approach the AKLT point. Moreover, the correlation length obtained at the AKLT point was found to be exactly $\frac{1}{\log(3)}$  and deviating slightly in its vicinity. This again matches the behavior obtained for the energy gaps and manifests the characteristic length scale associated with the physical model.  As a future work, we wish to extend our analysis for fermionic SPT phases and to deepen the relation
between the topological phases and CFT.

\section*{Acknowledgements}
This work has also been financed by the Spanish grants PGC2018-095862-B-C21, PGC2018-094763-B-I00, PID2019-105182GB-I00, QUITEMAD+ S2013/ICE-2801, SEV-2016-0597 of the ``Centro de Excelencia Severo Ochoa" Programme and the CSIC Research Platform on Quantum Technologies PTI-001.
\begin{widetext}
\appendix

\newpage
\section{Entanglement spectrum of AKLT state}
\label{AppendixA}
\subsection{Contiguous case}
We start with the decomposition of the AKLT state, as given in Eq. (\ref{eqn:AKLT_MPS}) of the main text,
\begin{eqnarray}
|\psi\rangle_{\text{AKLT}}&=&\sum_{i_1 i_2 \dots i_N} \text{Tr}(A_{i_1} A_{i_2}\dots A_{i_N}) |i_1 i_2\dots i_N\rangle.
\label{eqn:AKLT_MPS_A}
\end{eqnarray}
Using  the basis vectors $\{|\alpha\rangle\}$ of the $A_i$ matrices the above equation can be equivalently written as  
\begin{eqnarray}
|\psi\rangle_{\text{AKLT}}&=&\sum_{\alpha}\sum_{i_1 i_2 \dots i_N} \langle \alpha|A_{i_1} A_{i_2} \dots A_{i_N}|\alpha\rangle  |i_1 i_2\dots i_N\rangle. \nonumber
\end{eqnarray}
Now for a bipartition $l:N-l$, we can split the state as follows
\begin{eqnarray}
|\psi\rangle_{\text{AKLT}}&=&\sum_{\alpha, \beta}\sum_{i_1 i_2 \dots i_N} \langle \alpha|A_{i_1} A_{i_2}\dots A_{i_{l}}|\beta\rangle \langle \beta|A_{i_{l+1}} \dots A_{i_N}|\alpha\rangle  |i_1 i_2\dots i_N\rangle. \nonumber\\
|\psi\rangle_{\text{AKLT}}&=&\sum_{\alpha, \beta} |\phi^l_{\alpha \beta}\rangle|\phi^{N-l}_{ \beta\alpha}\rangle,
\label{eqn:main}
\end{eqnarray}
where 
\begin{eqnarray}
|\phi_{\alpha \beta}^l\rangle&=&\sum_{i_1 i_2 \dots i_{l}} \langle \alpha|A_{i_1}A_{i_2}\dots A_{i_l} |\beta\rangle |i_1 i_2\dots i_{l}\rangle,\nonumber\\
|\phi_{ \beta \alpha}^{N-l}\rangle&=&\sum_{i_{l+1} i_{l+2} \dots i_{N}} \langle \beta| A_{i_{l+1}}A_{i_{l+2}}\dots A_{i_N} |\alpha\rangle |i_{l+1} i_{l+2}\dots i_{N}\rangle.
\label{expnasion}
\end{eqnarray}
Now for the choices of $A_i$ matrices $A_0=\sqrt{\frac{2}{3}}\sigma^+$, $A_1=-\sqrt{\frac{1}{3}}\sigma
_z$ and $A_2=-\sqrt{\frac{2}{3}}\sigma^-$, let us write  the eigenvectors of the transfer matrix  $E=\sum_{i=0}^{D^2-1} \gamma_i |R_i\rangle\langle L_i|$  again 
\begin{eqnarray}
|R_0\rangle=|L_0\rangle&=&\frac{1}{\sqrt{2}} (|00\rangle +|11\rangle),\nonumber\\
|R_1\rangle=|L_1\rangle&=&\frac{1}{\sqrt{2}} (|00\rangle -|11\rangle),\nonumber\\
|R_2\rangle=|L_2\rangle&=&|01\rangle,\nonumber\\
|R_3\rangle=|L_3\rangle&=&|10\rangle,
\label{transfer_vec1}
\end{eqnarray}
with $\gamma_0=1, \gamma_1=\gamma_2=\gamma_3=\gamma=-\frac{1}{3}$.\

If we now use the $|R_i\rangle$ matrices as expressed in Eq. (\ref{transfer_vec1}), we can show that except $|\phi _{00}\rangle$ and $|\phi_{11}\rangle$ all other pairs of $|\phi_{\alpha\beta}\rangle$ are mutually orthogonal. This can be seen as follows.
\begin{eqnarray}
\langle \phi_{\alpha\beta}^l|\phi_{\alpha'\beta'}^l\rangle&=&\sum_{i_1 i_2\dots i_{l}} \langle \alpha |A^*_{i_1} A^*_{i_2}\dots A^*_{i_l}|\beta\rangle  \langle \alpha'|A_{i_1} A_{i_2} A_{i_3}\dots A_{i_l}|\beta'\rangle,\nonumber\\
&=&\sum_{i_1 i_2\dots i_{l}} \langle \alpha'|A_{i_1} A_{i_2} A_{i_3}\dots A_{i_l}|\beta'\rangle \langle \alpha |A^*_{i_1} A^*_{i_2}\dots A^*_{i_l}|\beta\rangle,\nonumber\\
&=&\sum_{i_1 i_2\dots i_{l}} \langle \alpha' \alpha | (A_{i_1}\otimes A_{i_1}^*) (A_{i_2}\otimes A_{i_2}^* )\dots (A_{i_l}\otimes A_{i_l}^*)|\beta' \beta\rangle,\nonumber\\
&=&\langle \alpha' \alpha|E^{l}|\beta' \beta\rangle.
\label{eqn_overlap}
\end{eqnarray}

Now expanding $E$ in terms of its right and left eigenvectors, we get

\begin{eqnarray}
\langle \phi_{\alpha\beta}^l|\phi_{\alpha'\beta'}^l\rangle&=&\langle \alpha'  \alpha |\left(\sum_k \gamma_k^l |R_k\rangle\langle L_k|\right)|\beta' \beta\rangle.
\end{eqnarray}

Let us now take the following overlaps

 \begin{eqnarray}
\langle \phi_{00}^l|\phi_{00}^l\rangle&=&\langle 00|E^l|00\rangle=\langle00|R_0\rangle \langle L_0|00\rangle+\gamma^l \langle 00|R_1\rangle \langle L_1|00\rangle=\frac{1}{2}(1+\gamma^l),\nonumber\\
\langle \phi_{00}^l|\phi_{01}^l\rangle&=&\langle 00|E^l|10\rangle=0,\nonumber\\
\langle \phi_{00}^l|\phi_{10}^l\rangle&=&\langle 00|E^l|01\rangle=0,\nonumber\\
\langle \phi_{00}^l|\phi_{11}^l\rangle&=&\langle 10|E^l|10\rangle=\langle 10|R_3\rangle \langle R_3|10\rangle=\gamma^l,\nonumber\\
\langle \phi_{01}^l|\phi_{00}^l\rangle&=&\langle 00|E^l|01\rangle=0,\nonumber\\
\langle \phi_{01}^l|\phi_{01}^l\rangle&=&\langle 00|E^l|11\rangle=\langle00|R_0\rangle \langle L_0|11\rangle+\gamma^l \langle 00|R_1\rangle \langle L_1|11\rangle=\frac{1}{2}(1-\gamma^l),\nonumber\\
\langle \phi_{01}^l|\phi_{10}^l\rangle&=&\langle 10|E^l|01\rangle=0,\nonumber\\
\langle \phi_{01}^l|\phi_{11}^l\rangle&=&\langle 10|E^l|11\rangle=0,\nonumber\\
\langle \phi_{10}^l|\phi_{00}^l\rangle&=&\langle 01|E^l|00\rangle=0,\nonumber\\
\langle \phi_{10}^l|\phi_{01}^l\rangle&=&\langle 01|E^l|10\rangle=0,\nonumber\\
 \langle \phi_{10}^l|\phi_{10}^l\rangle&=&\langle 11|E^l|00\rangle=\langle11|R_0\rangle \langle L_0|00\rangle+\gamma^l \langle 11|R_1\rangle \langle L_1|00\rangle=\frac{1}{2}(1-\gamma^l),\nonumber\\
\langle \phi_{10}^l|\phi_{11}^l\rangle&=&\langle 11|E^l|10\rangle=0,\nonumber\\
\langle \phi_{11}^l|\phi_{00}^l\rangle&=&\langle 01|E^l|01\rangle=\langle 01|R_2\rangle \langle R_2|01\rangle=\gamma^l.\nonumber\\
\langle \phi_{11}^l|\phi_{01}^l\rangle&=&\langle 01|E^l|11\rangle=0,\nonumber\\
\langle \phi_{11}^L|\phi_{10}^l\rangle&=&\langle 11|E^l|01\rangle=0,\nonumber\\
\langle \phi_{11}^l|\phi_{11}^l\rangle&=&\langle 11|E^l|11\rangle=\langle11|R_0\rangle \langle L_0|11\rangle+\gamma^l \langle 11|R_1\rangle \langle L_1|11\rangle=\frac{1}{2}(1+\gamma^l).
 \end{eqnarray}
Hence, from the above overlaps we can see $|\phi_{01}\rangle$ and $|\phi_{10}\rangle$ are orthogonal to each one of the basis states. Whereas, the basis states $|\phi_{00}\rangle$ and $|\phi_{11}\rangle$ are though orthogonal to $|\phi_{01}\rangle$ and $|\phi_{10}\rangle$, $\langle \phi_{00}|\phi_{11}\rangle$ or $\langle \phi_{11}|\phi_{00\rangle}$ has nonzero overlap. However,   if we take the following linear combination $|\phi_{\pm}\rangle=\frac{1}{\sqrt{2}} (|\phi_{00}\rangle\pm|\phi_{11}\rangle)$, then all these four states become mutually orthogonal.
\begin{eqnarray}
\langle \phi_+^l|\phi_-^l\rangle &=&\frac{1}{2} \left(\langle \phi_{00}^l|\phi_{00}^l\rangle-\langle \phi_{00}^l|\phi_{11}^l\rangle+\langle \phi_{11}^l|\phi_{00}^l\rangle_-\langle \phi_{11}^l|\phi_{11}^l\rangle\right), \nonumber\\
&=&\frac{1}{2}\left(\frac{1+\gamma^l}{2}-\gamma^l+\gamma^l-\frac{1+\gamma^l}{2}\right)=0.
\end{eqnarray}
\begin{eqnarray}
\langle \phi_+^l|\phi_+^l \rangle &=&\frac{1}{2}\left(\langle \phi_{00}^l|\phi_{00}^l\rangle+ \langle \phi_{00}^l|\phi_{11}^l\rangle+\langle \phi_{11}^l|\phi_{00}^l\rangle+\langle \phi_{11}^l|\phi_{11}^l\rangle\right), 
\nonumber\\
&=&\frac{1}{2}\left(\frac{1+\gamma^l}{2}+\gamma^l+\gamma^l+\frac{1+\gamma^l}{2}\right)=\frac{1+3\gamma^l}{2}.\nonumber\\
\langle \phi_-^l|\phi_-^l\rangle &=&\frac{1}{2}\left(\langle \phi_{00}^l |\phi_{00}^l\rangle-\langle \phi_{00}^l|\phi_{11}^l\rangle-\langle \phi_{11}^l |\phi_{00}^l\rangle_+ \langle \phi_{11}^l|\phi_{11}^l\rangle\right), \nonumber\\
&=&\frac{1}{2}\left(\frac{1+\gamma^l}{2}-\gamma^l-\gamma^l+\frac{1+\gamma^l}{2}\right)=\frac{1-\gamma^l}{2}.
\end{eqnarray}

 Therefore, we can re-write the expression of $|\psi\rangle_{AKLT}$ as follows
\begin{eqnarray}
|\psi\rangle_{AKLT}&=&\sum_{\alpha, \beta} |\phi^l_{\alpha\beta}\rangle |\phi_{\beta \alpha}^{N-l}\rangle,  \nonumber\\
&=&|\phi_{01}^l\rangle|\phi_{10}^{N-l}\rangle+|\phi_{10}^l\rangle|\phi_{01}^{N-l}\rangle+{|\phi_{+}^l\rangle}{|\phi}_{+}^{N-l}\rangle+{|\phi_{-}^l\rangle}{|\phi}_{-}^{N-l}\rangle,\nonumber\\
&=&\sqrt{\lambda_{01}^l}\sqrt{\lambda_{10}^{N-l}}\frac{|\phi_{01}^l\rangle}{\sqrt{\lambda_{01}^l}}\frac{|\phi_{10}^{N-l}\rangle}{\sqrt{\lambda_{10}^{N-l}}}+
\sqrt{\lambda_{10}^l}\sqrt{\lambda_{01}^{N-l}}\frac{|\phi_{10}^l\rangle}{\sqrt{\lambda_{10}^l}}\frac{|\phi_{01}^{N-l}\rangle}{\sqrt{\lambda_{01}^{N-l}}}
+\sqrt{{\lambda}^l_+}\sqrt{{\lambda}^{N-l}_+} \frac{ {|\phi}_{+}^l\rangle}{\sqrt{{\lambda}^l_+}} \frac{{|\phi}_{+}^{N-l}\rangle}{\sqrt{\lambda_{+}^{N-l}}}
+\sqrt{{\lambda}^l_-}\sqrt{{\lambda}^{N-l}_-}+ \frac{ {|\phi}_{-}^l\rangle}{\sqrt{{\lambda}^l_-}} \frac{{|\phi}_{-}^{N-l}\rangle}{\sqrt{{\lambda}^{N-l}_-}},\nonumber
\\
&=&\sqrt{\lambda_{01}^l}\sqrt{\lambda_{10}^{N-l}}|\tilde {\tilde \phi}_{01}^l\rangle|\tilde {\tilde \phi}_{10}^{N-l}\rangle+\sqrt{\lambda_{10}^l}\sqrt{\lambda_{01}^{N-l}}|\tilde{\tilde \phi}_{10}^l\rangle|\tilde {\tilde \phi}_{01}^{N-l}\rangle+\sqrt{{\lambda}^l_+}\sqrt{{\lambda}^{N-l}_+}{|\tilde {\tilde \phi}}_{+}^l\rangle{|\tilde{\tilde\phi}}_{+}^{N-l}\rangle+\sqrt{{\lambda}^l_-}\sqrt{{\lambda}^{N-l}_-}{|\tilde{\tilde\phi}}_{-}^l\rangle{|\tilde{\tilde \phi}}_{-}^{N-l}\rangle,
\end{eqnarray}
 with $|\tilde{\tilde\phi}\rangle$ the normalized states and
 \begin{eqnarray}
 \lambda^l_{01}&=&\langle \phi_{01}^l|\phi_{01}^l\rangle,\qquad~\lambda^{N-l}_{01}=\langle \phi_{01}^{N-l}|\phi_{01}^{N-l}\rangle,\nonumber\\
  \lambda^l_{10}&=& \langle \phi_{10}^l|\phi_{10}^l\rangle,\qquad~\lambda^{N-l}_{10}=\langle \phi_{10}^{N-l}|\phi_{10}^{N-l}\rangle,\nonumber\\
  \lambda^l_{+}&=& \langle \phi_{+}^l|\phi_{+}^l\rangle,\qquad~\lambda^{N-l}_{+}= \langle \phi_{+}^{N-l}|\phi_{+}^{N-l}\rangle,\nonumber\\
    \lambda^l_{-}&=&\langle \phi_-^l|\phi_{-}^l\rangle,\qquad~\lambda^{N-l}_{-}=\langle \phi_{-}^{N-l}|\phi_{-}^{N-l}\rangle.
\end{eqnarray}

 Hence, the four eigenvalues of the reduced state obtained for the $l: N-l$ bipartition is given by 
 \begin{eqnarray}
 \lambda_0&=&\lambda^l_+ \lambda^{N-l}_+= \frac{  (1+3\gamma^{l}) (1+3\gamma^{N-l})}{4},\nonumber\\
 \lambda_1&=&\lambda^l_- \lambda^{N-l}_-=\frac{(1-\gamma^{l})(1-\gamma^{N-l})}{4}, \nonumber\\
 \lambda_2&=&\lambda^l_{01} \lambda^{N-l}_{10}=\frac{(1-\gamma^{l})(1-\gamma^{N-l})}{4}, \nonumber \\
\lambda_3 &=& \lambda^l_{10} \lambda^{N-l}_{01}=\frac{(1-\gamma^{l})(1-\gamma^{N-l})}{4}.
 \end{eqnarray}
Now to normalize the eigenvalues, we divide the above  $\lambda_i$'s by $\sum_i \lambda_i=1+3\gamma^{N}$, and finally get 
\begin{eqnarray}
 \tilde{\lambda}_0&=& \frac{1}{4}  \frac{(1+3\gamma^{l}) (1+3\gamma^{N-l})}{1+3\gamma^N},\nonumber\\
 \tilde{\lambda}_1&=&\frac{1}{4}   \frac{(1-\gamma^{l})(1-\gamma^{N-l})}{1+3\gamma^N}, \nonumber\\
 \tilde{\lambda}_2&=& \frac{1}{4}  \frac{(1-\gamma^{l})(1-\gamma^{N-l})}{1+3\gamma^N}, \nonumber \\
\tilde{\lambda}_3 &=& \frac{1}{4}  \frac{(1-\gamma^{l})(1-\gamma^{N-l})}{1+3\gamma^N}.
 \end{eqnarray}

Along with this, here we present the transformation applied on the eigenbasis of the entanglement Hamiltonians which transforms $|\phi_{\alpha\beta}\rangle$ to $|\tilde{\phi}_{\alpha\beta}\rangle$, given by 
\begin{eqnarray}
|\tilde{\alpha}\rangle&=&|\alpha\rangle,\nonumber\\
 | \tilde{\beta}\rangle&=&e^{i\pi {(1-\beta)}}|{1-\beta}\rangle.\nonumber\\
 \label{eqn:Basis_transformation}
\end{eqnarray}

 \subsection{Non-contiguous case}
In case of non-contiguous bipartition of the state, say $A_1:B_1:A_2:B_2$ we can begin from the same decomposition, given by  
\begin{eqnarray}
|\psi\rangle_{\text{AKLT}}=&&\sum_{\alpha \beta} |\phi^{l'}_{\alpha\beta}\rangle|\phi_{ \beta\alpha}^{N-l'}\rangle,\nonumber
\end{eqnarray}
where $ |\phi^{l'}_{\alpha \beta}\rangle=\sum_{\gamma \delta}  |\phi_{\alpha \gamma}^{l_{A_1}}\rangle  |\phi_{\gamma \beta}^{l_{B_1}}\rangle$ and $|\phi_{ \beta,\alpha}^{N-l'}\rangle=\sum_{\gamma \delta}|\phi^{l_{A_2}}_{\beta \delta}\rangle |\phi^{l_{B_2}}_{\delta \alpha}\rangle.$ Now plugging these in the above equation and bringing the A-part ($A_1 \cup A_2$) and B-part ($B_1\cup B_2$) together, we get

\begin{eqnarray}
|\psi\rangle_{\text{AKLT}}&=& \sum_{\alpha  \gamma  \beta \delta}  |\phi_{\alpha \gamma}^{l_{A_1}}\rangle  |\phi_{\gamma \beta}^{l_{B_1}}\rangle |\phi^{l_{A_2}}_{\beta \delta}\rangle |\phi^{l_{B_2}}_{\delta \alpha}\rangle,\nonumber\\
&=& \sum_{\alpha  \gamma  \beta \delta} \underbrace{ |\phi_{\alpha \gamma}^{l_{A_1}}\rangle  |\phi^{l_{A_2}}_{\beta \delta}}_{A}\rangle \underbrace{|\phi_{\gamma \beta}^{l_{B_1}} \rangle |\phi^{l_{B_2}}_{\delta \alpha}}_{B}\rangle.
\end{eqnarray}
Therefore, the above situation becomes exactly similar to the contiguous case and hence we can proceed like before and get the entanglement spectrum for the reduced subsystem $A_1 \cup A_2$ of the AKLT state. The eigenvalues of the reduced state with the condition $l_{A_1}=l_{A_2}=\ell_A$ and $l_{B_1}=l_{B_2}=\ell_B$ are given in Eq. (\ref{eqn:lambda_noncontiguous1}) of the main text. However, one can also consider one of the subsystems of the B-part is infinite, i.e. $l_{B_2}\rightarrow \infty$. In this case, the set of eigenvalues can be expressed as 

\begin{eqnarray}
\lambda_i&=&\frac{1}{16} (1-\gamma^{\ell_A})^2  (1-\gamma^{\ell_B}),~i=1,2,3,4,5, \nonumber\\
\lambda_j&=& \frac{1}{16} (1+3\gamma^{\ell_A}) (1-\gamma^{\ell_A}) (1-\gamma^{\ell_B}),~j=6, 7, 8,\nonumber \\
\lambda_k&=&\frac{1}{16}  \Big(\zeta_1+\sqrt{(\zeta_1^2-\zeta_2)}\Big),~k=9, 10, 11, \nonumber\\\lambda_l&=&\frac{1}{16}  \Big(\zeta_1-\sqrt{(\zeta_1^2-\zeta_2)}\Big),~l=12, 13, 14, \nonumber\\
\lambda_{15}&=& \frac{1}{16}  \Big(\frac{\zeta_3+\sqrt{(\zeta_3^2-4\zeta_4)}}{2}\Big),\nonumber\\
\lambda_{16}&=&\frac{1}{16}  \Big(\frac{\zeta_3-\sqrt{(\zeta_3^2-4\zeta_4)}}{2}\Big),
\end{eqnarray}
where
\begin{eqnarray}
\zeta_1&=&(1+\gamma^{\ell_A})(1-\gamma^{\ell_A})(1+\gamma^{\ell_B}),\nonumber\\
\zeta_2&=&(1+3\gamma^{\ell_A}) (1-\gamma^{\ell_A})^3 (1+3\gamma^{\ell_B})(1-\gamma^{\ell_B}),\nonumber\\
\zeta_3&=&(1+3\gamma^{\ell_A})^2+(1-\gamma^{\ell_A})^2 (1+2\gamma^{\ell_B}),\nonumber\\
\zeta_4&=&(1+3\gamma^{\ell_A})^2(1-\gamma^{\ell_A})^2(1+3\gamma^{\ell_B})(1-\gamma^{\ell_B}).\nonumber
\end{eqnarray}

\section{Analytical derivation of SOP at AKLT point}
\label{AppendixB}
In this section, we provide the results obtained once the operator $\widehat{S_z}$ acts on the set of vectors $|{R}_0\rangle, |\tilde{R}_0\rangle, |{L}_0\rangle, |\tilde{L}_0\rangle$, given by 
\begin{align}
\widehat{S_z}|R_0\rangle=\frac{2}{3}|\tilde{R}_0\rangle, \; \widehat{S_z}|R_1\rangle=-\frac{2}{3}|\tilde{R}_1\rangle, \; \widehat{S_z}|R_2\rangle=0, \;
\widehat{S_z}|R_3\rangle=0, \nonumber\\
\langle L_0|\widehat{S_z}|=-\frac{2}{3} \langle \tilde{R}_0|, \;
\langle L_1| \widehat{S_z}=\frac{2}{3} \langle\tilde{L}_1|, \;
\langle L_2| \widehat{S_z}=0, \;
\langle L_3| \widehat{S_z}=0, \nonumber\\
\widehat{S_z}|\tilde{R}_0\rangle=-\frac{2}{3}|R_0\rangle, \; \widehat{S_z}|\tilde{R}_1\rangle=\frac{2}{3}|R_1\rangle, \; \widehat{S_z}|\tilde{R}_2\rangle=0, \;
\widehat{S_z}|\tilde{R}_3\rangle=0, \nonumber\\
\langle \tilde{L}_0|\widehat{S_z}|=-\frac{2}{3} \langle R_0|, \;
\langle \tilde{L}_1| \widehat{S_z}=\frac{2}{3} \langle L_1|, \;
\langle \tilde{L}_2| \widehat{S_z}=0, \;
\langle\tilde{L}_3| \widehat{S_z}=0. 
\end{align}

In addition to this, we provide an alternative derivation of the independence of SOP on the inter-site distance $l$.  
 From Sec. \ref {sec:SOP} and the relations derived above one can  verify the following relations
 \beq
 \tilde{E} \hat{S}_z = \hat{S}_z E, 
 \label{eqn1}
 \eeq
where 
\beq
\tilde{E} = {e^{ - i \pi \widehat{S_z}}}. 
\eeq 
Eq. (\ref{eqn1}) shows that the string order parameter is independent on the size of the string
\beq
{\cal O}_{\theta}(\ell, N_O) = \text{Tr} ( E^{N_O- \ell-2} \hat{S_z} \tilde{E}^\ell \hat{S_z} ) = \text{Tr}( E^{N_O- \ell-2} \hat{S_z}  \hat{S_z} {E}^\ell) = 
\text{Tr} ( E^{N_O-2} \hat{S_z}  \hat{S_z}).
\label{eqn2}
\eeq
Hence, ${\cal O}(\ell, N_O)$ only depends on $N_O$ as given in Eq. (\ref{eqn:SOP_fitting}) in the main text
\beq
{\cal O}_{\theta_{AKLT}}(N_O) = \mathcal{O}_{\theta_{AKLT}}(\infty)+\mathcal{A}(\theta_{AKLT},\infty) \exp\(-\frac{N_O}{\xi_{\mathcal{O}}(\theta_{AKLT}, \infty)}\). 
\eeq

One can also verify that 
 \beq
 {E} \hat{S}_z   = \hat{S}_z \tilde{E}, 
 \eeq
which allows one to write Eq. (\ref{eqn2}) as
\beq
{\cal O}_{\theta}(\ell, N_O) = 
\text{Tr} ( \tilde{E}^{N_O-2} \widehat{\tilde{S}_z}  \widehat{\tilde{S}_z}).
\eeq
This eq. suggests that the AKLT state can also be written in terms of $\tilde{A}_k$ matrices for which $\tilde{E}$
becomes the transfer matrix. \

 Based on Eq. (\ref{eqn1}) we can propose a generalization of the string order parameter using the MPS formalism.
A string order parameter will be defined as
\beq
{\cal O}_{\theta}(\ell, N_O)  =  \langle {\cal S}_i  \prod_{j=i+1}^{i+ \ell} {\cal T}_j  \;  {\cal S}_{i+\ell+1} \rangle,
\eeq
where ${\cal S}_i$ and ${\cal T}_i$ are local operators such that for the corresponding MPS state
the dressed operators $\hat{{\cal S}}$ and $\hat{{\cal T}}$ satisfy
 \beq
 \widehat{\mathcal{T}} \widehat{\mathcal{S}} = \widehat{\mathcal{S}}  E.
 \eeq
This equation  becomes exact for the AKLT state and in the other cases will be an approximation.

\end{widetext}

\end{document}